\documentclass[nohyper]{JHEP3}
\usepackage{amsmath,amssymb}
\usepackage{graphicx}
\usepackage{cite}

\newcommand{\be}{\begin{equation}}
\newcommand{\ee}{\end{equation}}
\newcommand{\bea}{\begin{eqnarray}}
\newcommand{\eea}{\end{eqnarray}}
\newcommand{\ba}{\begin{array}}
\newcommand{\ea}{\end{array}}
\newcommand{\balg}{\begin{align}}
\newcommand{\ealg}{\end{align}}

\newcommand{\lsim}
{\raise0.3ex\hbox{$\;<$\kern-0.75em\raise-1.1ex\hbox{$\sim\;$}}}
\newcommand{\gsim}
{\raise0.3ex\hbox{$\;>$\kern-0.75em\raise-1.1ex\hbox{$\sim\;$}}}

\title{Coupled dark matter-dark energy in light of near universe observations}
\author{Laura Lopez Honorez$^{1}$, Beth A.~Reid$^{2}$, Olga Mena$^{3}$, Licia Verde$^{4,2}$ and 
Raul Jimenez$^{4,2}$.\\

$^{1}$ Physics Department and Instituto de Fisica Teorica UAM/CSIC, 28049 Cantoblanco, Madrid, Spain and\\
Service de Physique Th\'eorique, ULB, 1050 Brussels, Belgium\\
$^{2}$ Institute for Sciences of the Cosmos (ICC), University of Barcelona and IEEC, Barcelona 08028, Spain\\   
$^{3}$ Instituto de Fisica Corpuscular, IFIC, CSIC and Universidad de Valencia, Spain\\
$^{4}$ ICREA (Institucio Catalana de Recerca i Estudis Avan\c{c}ats)  }
\abstract{Cosmological  analysis based on currently available observations
are unable to rule out a sizeable coupling 
among the dark energy and dark matter fluids. We explore a variety of coupled 
dark matter-dark energy models, which satisfy cosmic microwave background constraints, in light of low redshift and near universe 
observations.  We illustrate the phenomenology of different classes of dark coupling models, paying particular attention in distinguishing between effects that appear only on the expansion history and those that appear in the growth of structure. We find that  while a broad class of dark coupling models are effectively models where general relativity (GR) is modified --and thus can be probed by a combination of tests for the expansion history and  the growth of structure--, there is a class of dark coupling models  where gravity is still GR, but the growth of perturbations is,  in principle  modified.  While this effect is small in the specific  models  we have considered, one should bear in mind that an inconsistency between reconstructed expansion history and growth may not uniquely indicate deviations from GR. Our low redshift constraints arise from cosmic velocities, redshift space distortions and dark matter abundance in galaxy voids. We find that  current data constrain  the dimensionless  coupling to be  $|\xi|<0.2$,  but prospects from forthcoming data are for a significant  improvement.  Future,  precise measurements of the Hubble constant, combined with
high-precision constraints on the growth of structure, could provide
the key to rule out dark coupling models which survive other tests. We
shall exploit as well weak equivalence principle violation arguments,
which have the potential to  highly disfavour a broad family of
coupled models. } 

\begin{document} 
\section{Introduction}

Cosmological probes~\cite{Komatsu:2008hk,Kowalski:2008ez,Tegmark:2006az,
Percival:2006gt,Reid:2009xm,Percival:2009xn,Komatsu:2010fb} indicate that the 
universe we observe today possesses a flat geometry and a mass energy density 
made of $\sim 30\%$ baryonic plus 
cold dark matter and $70\%$ dark energy, responsible for the late-time accelerated 
expansion. The most economical description of the cosmological measurements
attributes the dark energy to a Cosmological Constant (CC) in
Einstein's equations, representing an invariable vacuum energy
density, with constant equation of state $w=-1$. However, from the quantum field theory approach, the predicted energy for the vacuum fluctuations is
$\sim 120$ orders of magnitude larger than the observed value. This
situation is the so-called CC problem. In addition, there is no
proposal which explains naturally why the matter and the vacuum energy
densities give similar contributions to the universe's energy budget
at this moment in the cosmic history. This is the so-called \emph{why
  now?} problem, and a possible way to alleviate it is to assume a
time varying, dynamical fluid. The quintessence option consists of a
cosmic scalar \emph{quintessence} field  which
changes with time and varies across space, and it is slowly
approaching its ground state e.g.,~\cite{Peebles:1987ek,Ratra:1987rm,Wetterich:1994bg,Caldwell:1998je,Zlatev:1998tr,Wang:1999fa}. The quintessence equation of state is generally not constant through cosmic time. In principle, the quintessence field may couple to other fields. In practice, observations strongly constrain the
couplings to ordinary matter \cite{Carroll:1998zi}. In addition, due to the the dynamical nature of the quintessence field, any coupling to the baryons would lead to time variation of the constants of nature, which are being tightly constrained see e.g.,~\cite{Martins} and references therein.   However, interactions
within the dark sectors, i.e. between dark matter and dark energy, are
still allowed by observations.  A non-zero coupling in the dark sector could affect significantly the expansion history of the universe and the density perturbation evolution, changing the growth history of cosmological structures, see Refs.~\cite{Amendola:1999er,Amendola:1999dr,Amendola:1999qq,Amendola:2000uh,Amendola:2003wa,Valiviita:2008iv,He:2008si,Jackson:2009mz,Gavela:2009cy,CalderaCabral:2009ja,Valiviita:2009nu,Majerotto:2009np,Gavela:2010tm}. A number of studies have been devoted to analyze the constraints from Cosmic Microwave Background (CMB), large scale structure (LSS), Supernovae Ia  and Baryon Acoustic oscillations (BAO) on a variety of interacting models~\cite{Amendola:2006dg,Wang:2006qw,Guo:2007zk,Olivares:2007rt,Feng:2008fx,He:2008tn,He:2009pd,Gavela:2009cy,Valiviita:2009nu,Gavela:2010tm}.
Forecasts from CMB experiments, as the on going
Planck\footnote{http://www.sciops.esa.int/index.php?project=PLANCK}
and future satellite missions e.g.,\cite{Bock} on coupled cosmologies
have been recently addressed~\cite{Martinelli:2010rt}.

Coupled cosmologies, in order to satisfy CMB constraints, predict values for the cosmological parameters today which may differ substantially from the parameters values within non-interacting cosmologies. In order to fit high-precision CMB data available today, coupled cosmologies can \emph{hide} their effects at very low redshifts. Therefore, low redshift probes  are highly complementary and thus powerful to constrain interacting dark sector models. In this paper we focus on near-universe, low-redshift constraints in a variety of coupled dark matter-dark energy models.  We  explore the phenomenology of coupled models and consider what type of low-redshift observations are most suitable to improve present constraints. 
We pay attention  in distinguishing between effects that appear only on the expansion history and that can thus be tested with observations such as BAO and Supernova Ia and those that appear in the growth of structure.
In the spirit of Ref.~\cite{Koyama:2009gd} we shall exploit galaxy velocities and weak equivalence principle violation arguments as additional probes to tighten interactions among the dark sectors. We present as well unexplored, powerful constraints arising from dark matter abundance in voids, as well as from future, precise measurements of the Hubble constant $H_0$. Throughout this paper we assume a flat spatial geometry.

It was suggested that the dynamical equilibrium of collapsed
structures such as galaxy cluster could advocate in favour of an
interaction between dark matter and dark
energy~\cite{Bertolami:2007zm}. The idea is that the virial theorem is modified by the energy exchange between the dark
sectors leading to a bias in the estimation of the virial masses of
clusters when the usual virial conditions are employed. This provides an additional near universe probe of the  dark coupling. By comparing
weak lensing and X-ray mass-observables to the virial masses, the
authors of \cite{Abdalla2009107,Abdalla:2009mt} 
claim that available
data suggest that a small coupling could be present involving the
decay of dark energy  into dark matter. Notice that they also warn
the reader that the lack of knowledge about the errors arising from systematics
could weaken their conclusions.
 
In Section~\ref{sec:seci} we propose a classification of interacting cosmologies into two broad
families (\emph{DEvel} and  \emph{DMvel}) and, in each of these
families, two classes, depending on the scaling of the coupling with
the matter or dark energy densities. This section presents also the
background and the linear perturbation evolution for the different
cases and discuss instabilities. Section~\ref{lineargrowthblah} 
focuses on the phenomenology of these models compared to a $\Lambda$CDM model
and to uncoupled models with arbitrary equation of state parameter for
dark energy. 
Section~\ref{sec:obs} summarizes current constraints for interacting
models and  future prospects, devoting separate subsections to near 
universe $H_0$ measurements, skewness tests, dark matter
velocities  and void dark matter abundances. Finally, we draw our conclusions in
Section~\ref{sec:concl}.

\section{Two times two families of  dark coupling models}
\label{sec:seci}
In this section we present the background and linear perturbation
theory results for dark coupling models in general and then in the
context of specific forms of the  coupling.  The models that we
consider are naturally divided by  two features of the model: whether
the assumed momentum transfer is zero in the dark energy  rest frame
versus the dark matter rest frame. 
In each of  these cases the the energy transfer could be taken  to be 
proportional to $\rho_{\rm de}$ or to  $\rho_{\rm dm}$.


\subsection{Dark coupling models and modified gravity}
\label{sec:4moment-trsf}

At the level of the stress-energy tensor it is always possible to
introduce an interaction between the fluids of the dark sector in the following way~\cite{Kodama:1985bj}: 
\begin{eqnarray}
\nabla_\mu T^\mu_{({\rm dm})\nu} =Q_\nu \quad\mbox{and}\quad
\nabla_\mu T^\mu_{({\rm de})\nu} =-Q_\nu.
\label{eq:conservDMDE}
\end{eqnarray}
The  4-vector $Q_\nu$ governs the energy-momentum transfer between the dark
components and  $T^\mu_{({\rm dm})\nu}$ and $T^\mu_{({\rm de})\nu}$
are the energy-momentum tensors for the dark matter and dark energy fluids, respectively. 
Different expressions for the form of $Q_\nu$,  that arise
from a variety of motivations, can be found in  the literature.  Here
we attempt to classify the couplings in broad families, based in the
different phenomenology -from the astrophysical and cosmological point 
of view- they display.

We consider two families of four momentum-energy transfer $Q_\nu$. In the
first family of models (DEvel), the momentum exchange  $Q_\nu$ is
parallel to the dark energy four velocity\footnote{The  scale factor
  in the denominator is introduced because -at least at linear order-
  the velocity $u$ is $\propto a$, see Eq.~(\ref{eq:velproptoa}).} $u_{\nu}^{({\rm de})}$:
\begin{equation}
    Q_\nu = Q u_{\nu}^{({\rm de})}/a \qquad \mbox{(DEvel)} 
\label{eq:ue}
\end{equation} 
\noindent

In the second family of models (DMvel), $Q_\nu$ is parallel to the dark
matter four velocity~\cite{Valiviita:2008iv,He:2008si,Gavela:2009cy,Jackson:2009mz,Majerotto:2009np,Valiviita:2009nu,Koyama:2009gd} $u_{\nu}^{({\rm dm})}$:
\begin{equation}
    Q_\nu = Q u_{\nu}^{({\rm dm})}/a~ \qquad \mbox{(DMvel)}.
\label{eq:um}
\end{equation}

DEvel models $\propto \rho_{\rm dm}u_{\nu}^{({\rm de})}$ include  all quintessence coupled models, see for instance, Refs.~\cite{Damour:1990tw,Damour:1990eh, Wetterich:1994bg,Amendola:1999er,Zimdahl:2001ar,Farrar:2003uw,Das:2005yj,Zhang:2005jj,delCampo:2006vv,Bean:2007nx,Olivares:2007rt,Jackson:2009mz,Koyama:2009gd,Boehmer:2009tk}.
In DEvel models, there is no momentum transfer to the dark
energy frame, so that momentum must be conserved in the dark matter
frame.  This implies a fractional increase in the dark matter 
peculiar velocity equal and opposite to the fractional
change in energy density due to the presence of a coupling. 
This effect can be interpreted as an extra source of 
acceleration for the dark matter fluid, that
will appear clearly in the dark matter velocity perturbation equation,
see \S \ref{lineargrowthblah}.
By contrast, in DMvel models both momentum and
energy density are transferred from the dark matter system to the dark
energy one, and therefore the dark matter peculiar velocity field 
does not have this apparent force.

The extra force effect in DEvel models should not come as a surprise:
many quintessence coupled field models that appear
in the literature  can be  written as a scalar-tensor Brans-Dicke
\cite{BransDicke} gravity theory. Let us also recall that $f(R)$
gravity theories correspond to generalized
Brans Dicke (BD) theory with a BD parameter $w_{\rm BD} = 0$ or $=-3/2$, see
e.g. \cite{Sotiriou:2008rp} and references therein. The equations in the Einstein frame contain a new scalar field which satisfies Eqs.~(\ref{eq:conservDMDE}), being the energy momentum exchange $Q_\nu$ proportional to its 4-velocity (see e.g. \cite{DeFelice:2010aj}). 

The assumption of such a scalar interaction  (DEvel)
in the dark sector makes the acceleration of visible and dark
matter particles different, inducing a ``fifth force'' effect (only for
the dark matter), that is,  a violation of the equivalence principle,
for which the E$\ddot{\rm o}$tv$\ddot{\rm o}$s experiments constraints
do not apply.  As noted by Ref. \cite{AmendolaTocchiniVelentini},  a
large-scale fifth force in the dark sector might have substantial
effect  as a  mis-match in the relative distribution of baryons and
dark matter. 
DEvel models are therefore effectively ``modified gravity'' models. It
is well known that deviations from  the simpler $\Lambda$CDM  paradigm
in the form of general relativity (GR) modifications can be constrained following two approaches:
{\it a}) at the background evolution level as inconsistencies between
the high-redshift and the low-redshift universe
\cite{Ishaketal,AcquavivaVerde} and {\it b}) at the growth of
perturbation level: a modified gravity model with the same expansion
history as $\Lambda$CDM model, has a different growth of the dark
matter structures. This has been extensively discussed in
Refs.~\cite{jain,songpercival,guzzo:2008ac}. In fact, for some specific cases, there is a third possible
approach: {\it c}) using weak equivalence principle violation (WEPV) constraints.

In the following, we will see that option {\it a}) also
applies to DMvel models and we illustrate its potential in the
context of  those models in \S \ref{subsec:shifts}. This approach  only probes the background evolution and thus cannot be used alone to distinguish modified gravity  or dark coupling from a minimally coupled dark energy model  with general, time-dependent equation of state parameter $w(z)$. For what concerns
approach {\it b)}, we show in \S \ref{lineargrowthblah} that
the growth of perturbations in DMvel and DEvel coupled models  can differ
from the growth in an uncoupled model with identical background history. 
In the case of  DEvel models (which in practice are effectively modified 
gravity models) it is already very well known that growth can provide the key 
to break the existing degeneracy at the background level 
among GR and modified gravity. 
However, for DMvel models, the result we obtain here is a counterexample
to what is commonly accepted in the literature: DMvel models 
are not a modification of gravity, but their growth  can in principle still differ from 
the growth in a GR dynamical dark energy model which possesses the same 
background history  since $\rho_{\rm dm}(z)$ does not behave like dust (i.e., $\propto (1+z)^3$). See also \cite{Simpson:2010yt}. Finally, 
approach {\it c)} is specific of DEvel models (i.e., it is the
smoking gun of those models) and we will consider it quantitatively in \S \ref{sec:wepvnew}.

\subsection{Background evolution}
\label{sec:bck}
 In  Eq.(~\ref{eq:ue}) and (~\ref{eq:um}),  
$Q$ drives the energy exchange between dark
matter and dark energy. Indeed, one can easily show that for DEvel and DMvel families, the evolution
equations for the dark matter and dark energy background energy
densities reduce to: 
\begin{eqnarray}
  \label{eq:EOMm}
  \dot\rho_{\rm dm}+ 3\mathcal{H}\rho_{\rm dm} &=& Q\,,\\
\label{eq:EOMe}
 \dot\rho_{\rm de}+ 3 \mathcal{H}\rho_{\rm de}(1+ w)&=&- Q\,.
\end{eqnarray}
$\rho_{\rm dm} (\rho_{\rm de})$ denotes the dark matter (dark energy) energy density,  the dot indicates derivative with
respect to conformal time $d\tau = dt/a$, $\mathcal{H}= {\dot a}/a$
and  $w=P_{\rm de}/\rho_{\rm de}$ is the dark-energy equation of state ($P$
denotes the pressure).
We work with the Friedman-Robertson-Walker metric,
assuming a flat universe and  pressureless dark matter $w_{\rm dm} =
P_{\rm dm}/\rho_{\rm dm}=0$. 
The sign of $Q$ determines the direction of energy transfer. For positive $Q$, the energy flows from the dark energy system to dark matter one. For negative $Q$, the energy flow is reversed. Note that if $Q<0$, $\rho_{\rm dm}$ decreases with time because dark matter is being transformed into dark energy. 
 The presence of a coupling $Q$ also changes the dark
matter and dark energy redshift dependence acting as an extra
contribution to their effective equation of state.
  Indeed, the effective background equation of state for the two fluids are \cite{Majerotto:2009np,Gavela:2009cy}
\begin{equation}
w_{\rm dm}^{\rm eff}=-\frac{Q}{3{\cal H}\rho_{\rm dm}}\,;\,\,\,\,\, w_{\rm de}^{\rm eff}=w+\frac{Q}{3{\cal H}\rho_{\rm de}}~.
\end{equation}
Therefore e.g., a negative  $Q$ yields an effective equation of state for dark energy $w_{\rm de}^{\rm eff}$ that is more negative than $w \equiv P_{\rm de}/\rho_{\rm de}$. A negative $Q$ will also contribute as a positive pressure in the dark matter background equation. Note however that the
deceleration parameter satisfies, regardless of the presence of
non-zero dark coupling, 
\begin{equation}
  \label{eq:decel}
q=-  \frac{\dot {\mathcal H} }{{\mathcal H}^2}=\frac12 (1+3\,w\,\Omega_{\rm de})~,
\end{equation}
 where $\Omega_{\rm de}$ is the time dependent relative dark  energy
density. Therefore, we would still require $w < -1/3$ to have a universe
with accelerated expansion.

Because of the unknown nature of the dark sector,  to-date there is no
prescription in  fundamental theory  for a physically-motivated model
for the coupling between  the dark matter and dark energy fluids.  The
interaction term $Q$ is currently  mostly chosen in a phenomenological
way.  For the models considered in the literature in which the dark
coupling depends linearly on the dark sector energy densities, 
    for each of the two families DEvel and DMvel, we propose the
    definition of  two sub-classes of models:
\begin{eqnarray}
  Q&=&\Sigma\rho_{\rm dm} \qquad\mbox{(class I)} \label{eq:rhom}\\
 Q&=&\Sigma\rho_{\rm de} \qquad\mbox{(class II)}\label{eq:rhoe}
\end{eqnarray}
where $\Sigma$ denotes the interaction rate.  The former classification 
should be considered as a ``basis set'' for the study of coupled
models. For example, the coupling
$Q=\Sigma_1 \rho_{\rm dm}+\Sigma_2\rho_{\rm de}$ is also present in the
literature~\cite{He:2008si,Abdalla:2009mt,Abdalla2009107,Feng:2008fx,Wang:2006qw} and it does not strictly belongs to
class I or class II. We trust however that our analysis can guide the
reader to determine what would be the relevant test to apply
depending on the contribution that dominates at the scale or redshift
corresponding to the probes under investigation. To gain physical insight
on the phenomenology of coupled models we concentrate here on the simple
cases of ``pure'' class I or class II models in both DMvel or DEvel scenarios.

In coupled quintessence
models $\Sigma$ is proportional to the time derivative of the
quintessence field, see e.g. Refs.~\cite{Damour:1990tw,Damour:1990eh,
  Wetterich:1994bg,Amendola:1999er,Zimdahl:2001ar,Farrar:2003uw,Das:2005yj,Zhang:2005jj,delCampo:2006vv,Bean:2007nx,Olivares:2007rt,Jackson:2009mz,Koyama:2009gd,Boehmer:2009tk}. More phenomenologically motivated models have taken
$\Sigma$ to be proportional to the Hubble expansion rate, since the former 
cosmological parameter has the appropriate time dependence~\footnote{For
more discussion on a possible physical interpretation of this choice, see
e.g., \cite{Valiviita:2008iv,He:2008si}.}. Even if the 
$\Sigma\propto \cal H$ choice~\cite{Valiviita:2008iv,
  He:2008si,Gavela:2009cy,Jackson:2009mz}, is much more easy to handle mathematically than the $\Sigma\propto  H_0$ case~\cite{Valiviita:2008iv,Majerotto:2009np} (being $H_0$ the present-day value of the Hubble parameter), we will consider both possibilities in this paper.
  
It is important to keep in mind that the expansion history  does not
depend on the choice DEvel or DMvel (but  does depend on the form of
$Q$).  Unfortunately, an analytic form for the expansion history $H(z)$
cannot be written for generic $Q$. The only case where it can be done
is in the case of $Q=\xi {\cal H}\rho_{\rm de, dm}$ (a
mathematically easy to handle coupling) where both the coupling $\xi$  and the equation of state parameter $w$ are constant (see Eqs.~(\ref{eq:hz}) and (\ref{eq:hz2}) in Appendix A and e.g.,\cite{Gavela:2009cy}). Even if DMvel and DEvel models provide the same background
history,  the perturbation evolution is dramatically different, as we
shall see in the next subsection.  Therefore, geometrical probes   alone are
unable to distinguish among the two of them --or distinguish dark coupling from a uncoupled dark energy model with a $w(z)$--, even if, as we will see
in \S \ref{lineargrowthblah}, the two models are fundamentally different  and different from minimally coupled dark energy models.

\subsection{Linear perturbations}
\label{sec:lin}
In the Newtonian gauge, the perturbed FRW metric at linear order in scalar
perturbations is given by: 
\begin{equation}
  ds^{2}=a^2[-(1+2\Psi )d\tau^{2}+(1-2\Phi )dx_{i}dx^{i}]~,
\label{eq:metric}
\end{equation}
\noindent
where $a$ is the scale factor, $\tau$ is the conformal time, $x^i$ is
the comoving coordinate, and  $\Psi$ and $\Phi$ are the scalar metric
perturbations in the Newtonian gauge.

The four velocity of a fluid reads
\begin{equation}
  u_\nu=a(-(1+\Psi), v_i)~,
  \label{eq:velproptoa}
\end{equation}
where $v_i$ is the fluid component peculiar velocity.   
Since baryons are not coupled to the dark energy fluid, the continuity and the Euler equations for the baryons after decoupling are equivalent to those in uncoupled cosmologies:
  \begin{eqnarray}
\label{eq:deltab}
\dot\delta_{\rm b}  & = & -\theta_{\rm b}+3\dot \Phi\,~; \\
 \label{eq:thetab}
 \dot \theta_{\rm b}  & = & -{\mathcal H}\theta_{\rm b}+ 
 k^2 \Psi\,~,
\end{eqnarray}
where $\delta \equiv \delta \rho /\rho$ is the fluid energy density perturbation and $\theta \equiv \partial_i v^i$ is the divergence of the fluid proper velocity $v^i$. We have assumed that baryons behave as a barotropic fluid with
$dP_b/d\rho_b= \delta P_b/\delta \rho_b=0$, and we will assume the same
properties for dark matter.
For the coupled dark matter and dark energy components, we obtain, at linear order:
 \begin{eqnarray}
\label{eq:deltambe}
\dot\delta_{\rm dm}  & = & -(\theta_{\rm dm}-3\dot \Phi)
+\frac{Q}{\rho_{\rm dm}} \left(\frac{\delta Q}{Q}-\delta_{\rm dm}+\Psi \right)\, \\
 \label{eq:thetames}
 \dot \theta_{\rm dm}  & = & -{\mathcal H}\theta_{\rm dm} 
+(1-b)\frac{Q}{\rho_{\rm dm}}(\theta_{\rm de}-\theta_{\rm dm})+ k^2 \Psi\,\\  
\dot\delta_{\rm de}  & = & -(1+w)(\theta_{\rm de}-3\dot \Phi)
-\frac{Q}{\rho_{\rm de}} \left(\frac{\delta Q}{Q}-\delta_{\rm de}+\Psi
\right)\nonumber\\
& &-3 {\mathcal H}\left(\hat c_{s\,{\rm de}}^2 -w\right)\left[ \delta_{\rm de} +
{\mathcal H} \left( 3(1+w) +\frac{Q}{\rho_{\rm de}} \right)\frac{\theta_{\rm de}}{k^2}
\right]~ \label{eq:deltaees}\\  \nonumber\\
\dot \theta_{\rm de}  & = & -{\mathcal H}\left(1-3\hat c_{s\,{\rm de}}^2 -\frac{\hat
  c_{s\,{\rm de}}^2+b}{1+w} \frac{Q}{{\mathcal H}\rho_{\rm de}}
\right)\theta_{\rm de}+\frac{k^2}{ 1+w}\hat c_{s\,{\rm de}}^2 \delta_{\rm de}+ k^2 \Psi-b\,\frac{Q}{\rho_{\rm de}}\frac{ \theta_{\rm dm}}{1+w} \,.\nonumber\\\label{eq:thetaees}
\end{eqnarray}
We use the $b$ notation introduced in  Ref.~\cite{Jackson:2009mz},
where $b=0$ refers to DEvel models with $Q_\nu\propto u_\nu^{({\rm
    de})}$ (\ref{eq:um}) and $b=1$ refers to DMvel models  with $Q_\nu\propto
u_\nu^{({\rm dm})}$ (\ref{eq:ue}); $\hat c_{s\,{\rm de}}^2$ is
the dark energy pressure perturbation sound speed in the rest frame of
dark energy.
For the derivation of  Eq.~(\ref{eq:deltaees}), we have used the following relation: 
\begin{eqnarray}
{\delta P_{\rm de}}&=&\hat c_{s\,{\rm de}}^2 \delta \rho_{\rm de} +(\hat c_{s\,{\rm de}}^2-
c_{a\,{\rm de}}^2) \dot \rho_{\rm de} \frac{\theta_{\rm de}}{k^2}\,,
 \label{eq:dpcs}
\end{eqnarray}
which is valid in both Newtonian and synchronous gauges. If not otherwise
stated, in the following, we assume $\hat c_{s\,{\rm de}}^2=1$ (as for
quintessence)\footnote{Notice  that current data still allows for
  $c_s^2\sim 0$. However, in Ref.~\cite{Ballesteros:2010ks}, it is
  shown that the two limits $\hat{c}_s^2 \sim 0$ and $\hat{c}_s^2~1$
  could be discriminated by future experiments.}. The dark energy  density
perturbations will not cluster significantly if the  sound speed for
the dark energy $\hat c_{s\,{\rm de}}^2=1$, and therefore, in the
following, we can safely neglect dark energy  perturbations in the
perturbation evolution  (see also the discussion in Refs.~\cite{CalderaCabral:2009ja,Koyama:2009gd}).

Notice that the Euler equation for the dark matter fluid is \emph{only} 
modified in the first family of models (DEvel) considered here, where 
$Q_\nu \propto u_\nu^{({\rm de})}$ case, violating therefore the 
weak equivalence principle, as we shall see in \S
\ref{sec:wepvnew}. This feature of DEvel models was previously discussed in \S \ref{sec:4moment-trsf}.

The growth equations  for dark matter and baryons can easily be derived
from Eqs.~(\ref{eq:deltab}), (\ref{eq:thetab}),~(\ref{eq:deltambe})
and~(\ref{eq:thetames}) going to the Newtonian limit (i.e. for $k\ll
\cal H$). The growth equation for dark matter is however rather
sensible to the type of coupling. 

The simplest case is the one of DMvel $Q \propto \rho_{\rm dm}$ (class I) coupled models for which we recover the standard perturbation equations at linear order as well as the  growth equation  for the two matter
fluids $\alpha=\rm{dm, b}$:
\begin{equation}
\delta_{\alpha}''=-  (2-q)\frac{\delta_{\alpha}'}{a}+ \frac32 \left(\Omega_{\rm dm} 
  \frac{\delta_{\rm dm}}{a^2}+ \Omega_{\rm b}  \frac{\delta_{\rm b}}{a^2}\right)~,
  \label{eq:eqf2} 
\end{equation}
where $q$ is the deceleration parameter of Eq.~(\ref{eq:decel})  and $'\equiv d/da$. 
Notice that the evolution equations for baryons are always the standard
ones because they are not affected by the coupling introduced in Eq.~(\ref{eq:conservDMDE}). For the dark matter perturbations in DMvel  $Q \propto \rho_{\rm dm}$  models, the
difference with non-interacting cosmologies arise exclusively due to
the different background evolution of the quantities ${\mathcal H}$
and $\Omega_{\rm dm}$~\cite{CalderaCabral:2009ja,Koyama:2009gd}. The growth equation is not modified  but the scaling with redshift of $\rho_{\rm dm}$ is different from that of a conserved pressureless fluid. As a result
 these models are effectively indistinguishable from minimally coupled dark energy models with a $w(z)$.

For all the other cases, i.e.  DMvel $Q\propto \rho_{\rm de}$ class II and  DEvel class I $\&$ II,
the dark matter linear perturbation equations and the growth equations are
 both modified by the presence of a coupling term $Q$. 
In order to derive the growth equation, an explicit form for the
interaction term $Q$ must be assumed, as we shall see in \S \ref{lineargrowthblah}.

\subsection{Instabilities}
The existence of non adiabatic, early time instabilities in coupled
models is a well known
phenomena~\cite{Valiviita:2008iv,He:2008si,Chongchitnan:2008ry,Gavela:2009cy,Jackson:2009mz,Corasaniti:2008kx,Majerotto:2009np}. In
brief, the dark coupling term which appears  through
$\dot \rho_{de}$ in the dark energy pressure perturbations of Eq.~(\ref{eq:dpcs}) is a
source  for early time instabilities at large scales in the dark sector. 
The latter could arise in coupled models when the dark coupling term
dominate  on the standard (uncoupled) non adiabatic contributions to the
dark energy pressure perturbation. The instability is then rapidly
transferred to the others fluids components and to the curvature
perturbation related to them, producing a non viable
cosmological scenario.    

Here we briefly review the instability issue and provide a general
 recipe to avoid instabilities in the perturbation evolution in the
 case of a constant equation of state $w$ ($w_{\rm de}^{\rm eff}$ can
 still vary).  This recipe will be relevant throughout the paper, where 
 the range and sign for the dimensionless parameters describing 
the coupling have been chosen to ensure an instability-free perturbation 
evolution. 
 
The onset of non-adiabatic instabilities depends on the form of the dark 
coupling $Q$ (class I or class II), on the dark energy equation of state 
$w$ and on the $Q_{\nu}$ 4-velocity dependence (DMvel or DEvel).
We shall define a \emph{doom factor} ${\bf d}$ as \cite{Gavela:2009cy}  which is defined independently of the explicit form of the coupling $Q$:
\begin{equation}
 {\bf d} \equiv \frac{Q}{3\mathcal H\rho_{\rm de}(1+w)}\,.
\label{eq:maldito}
\end{equation}
If $|{\bf d}|>1$, the interaction among the two dark sectors drives the non-adiabatic contribution to the dark energy pressure perturbation. At large scales, or equivalently, at early times (${\cal H}/k\gg 1$), the leading contributions in $Q$, or equivalently in ${\bf d}$, to the second order differential equation for $\delta_{\rm de}$ reads:
 \begin{eqnarray}
  \label{eq:grstrongfullours}
\delta_{\rm de}''&\simeq&  \,3\,{\bf d}\,(\hat c_{s\,{\rm de}}^2+b)\left(\,\frac{\delta_{\rm de}'}{a} 
    \,+\,3b\frac{\delta_{\rm de}}{a^2}\frac{(\hat c_{s\,{\rm de}}^2-w)}{\hat c_{s\,{\rm de}}^2+b}
     \,+\,\frac {3(1+w)}{a^2}\delta[\,{\bf d}\,]\,\right)+...
  \end{eqnarray}
where $b=0$ ($b=1$) applies to DEvel (DMvel) models and $'\equiv d/da$. The sign of the coefficient of $\delta_{\rm de}^\prime$  in Eq.~(\ref{eq:grstrongfullours}) is, for $\hat c_{s\, {\rm de}}^2>0$, equivalent to the sign of the doom factor ${\bf d}$. A positive doom factor ${\bf d}>1$ can lead to large-scale instabilities. Therefore, a simple recipe to avoid non adiabatic instabilities is to
consider as viable models only those in which the energy transfer among the
dark sectors $Q$ and the (constant) equation of state of the dark energy
component $w$ are such that the doom  factor ${\bf d}$,
Eq.~(\ref{eq:maldito}), is always negative. 

Having defined all the tools, we briefly review the instability issue 
for the coupled models that will be analyzed in the following sections~\footnote{For a detailed treatment of the instability issue in each of the cases 
reviewed below, we refer the reader to Refs. \cite{Valiviita:2008iv,He:2008si,Chongchitnan:2008ry,Gavela:2009cy,Jackson:2009mz,Corasaniti:2008kx,Majerotto:2009np}}. We have shown that the key parameter is the doom factor, 
which only depends on background quantities such as $Q$. 
This means that the stability of the perturbations 
strongly depends on whether the model is a class I or a class II coupled model. However, the stability of the perturbations shows only a mildly dependence on the DEvel-DMvel classification of the model (see the $b$ dependence of
Eq.~(\ref{eq:grstrongfullours})).

For class II models, it is easy to verify that, for $Q=\xi {\cal H} \rho_{\rm de}$ with $\xi$ constant and negative, and $w>-1$, the doom factor is negative. In the following, we shall consider
 \begin{equation}
    Q_\nu= \xi {\mathcal H} \rho_{\rm de} u_{\nu}^{\rm dm}/a \qquad\mbox{and}\qquad  Q_\nu= \xi {\mathcal H} \rho_{\rm de} u_{\nu}^{\rm de}/a
\label{eq:ourm}
\end{equation}
with negative dimensionless coupling $\xi$ and $w>-1$. The two models above 
are, respectively, the DMvel class II and DEvel class II models 
 which we will analyze  next.

 It was shown in Ref.~\cite{Gavela:2009cy} that
for $Q=\xi {\cal H} \rho_{\rm dm}$,  the doom factor is always positive
 assuming $w>-1$ . It will also be enhanced by the $\rho_{\rm
   dm}/\rho_{\rm de}$ factor,  see Eq.~(\ref{eq:maldito}). A not very attractive solution to solve the
problem would be to consider a constant $w<-1$\cite{He:2008si,Gavela:2009cy,Jackson:2009mz}. Another avenue is to use a time dependent equation of state for dark energy. In Ref.\cite{Majerotto:2009np} it was shown that the  DMvel model
\begin{eqnarray}
    Q_\nu&=&- a\Gamma \rho_{\rm dm} u_{\nu}^{({\rm dm})}~,\label{eq:maart}
 \ \ \mbox{with} \quad w(a)=w_0\, a+w_e\,(1-a)~,
  \end{eqnarray}
is stable for a specific range of constant $w_0$, $w_e$ and
$\Gamma/H_0$. This is the DMvel class I model that we shall analyze.

In the case of DEvel class I models, we shall refer to the case of 
a coupled quintessence model:
\begin{equation}
  Q_{\nu}= \beta \rho_{\rm dm}\nabla_\nu \phi/M_p
\label{kk}
\end{equation}
where $\phi$ is the dark energy field driving the interaction,
$M_p$ denotes the Plank mass ($M_p=1/\sqrt{8\pi G_N}$) and $\beta$ will be considered to be a constant. Note that in this case $u_{\nu}^{({\rm de})}\propto \nabla_\nu \phi/\dot \phi$. 

The early time non adiabatic instabilities discussed here
are different from the \emph{adiabatic instabilities}. These adiabatic instabilities may arise at late times in quintessence coupled models if the sound speed of the total fluid gets negative~\cite{Bean:2007ny}, even if Ref.~\cite{Corasaniti:2008kx} has shown that the slow-rolling of the quintessence field may avoid such instabilities in some cases. In the following, the term instabilities will refer exclusively to non adiabatic instabilities.

\section{Distinguishing dark coupling models from arbitrary uncoupled $w(z)$ dark energy models}

\label{lineargrowthblah}

 At the background level, a dynamical, redshift-dependent,  effective equation of state $\tilde w(z)$, can be mimicked by the combination of a constant equation of state $w$ plus a dark matter-dark energy coupling $Q$~\cite{Gavela:2009cy}. Assume that the true, underlying cosmology possesses an interacting
dark energy fluid with constant, non phantom, equation of state $w>-1$. If  the background is analyzed setting the coupling to zero, one would reconstruct an effective  redshift dependent $\tilde w(z)$:

\begin{equation}
 \label{eq:wH}
\tilde w(z)=\frac13 \;\frac{ (1+z) dR_H/dz -3 R_H  }{R_H -\tilde{\Omega}^{(0)}_{\rm m} (1+z)^3 }\, ,
\end{equation}
where $R_H(z)$ is a function of the background of the coupled model,
\begin{equation}
  \label{eq:RHwz}
R_H(z)=\frac{H^2(z)}{H_0^2}~,
\end{equation}
and $\tilde{\Omega}^{(0)}_{\rm m}$ is   matter energy
density (dark matter plus baryons) as estimated from the CMB and extrapolated to $z=0$ assuming conserved pressureless dust.  An analogous relation between the redshift-dependent equation
of state and the luminosity distance and its derivative was firstly
presented in Ref.~\cite{Clarkson:2007bc}. 

In the following, we study uncoupled  models with  $\tilde w(z)$ versus interacting DMvel and DEvel class II cosmologies with $Q=\xi {\cal H} \rho_{\rm de}$.  We investigate how to distinguish among them, even if the background evolution in these cosmologies
is identical. For this particular type of coupling, $H(z)$ is given by Eq.~(\ref{eq:hz}),
where we consider $\xi < 0$ to 
avoid early-time instabilities.


\begin{figure}[h]
\begin{center}
\includegraphics[height=.4\textheight]{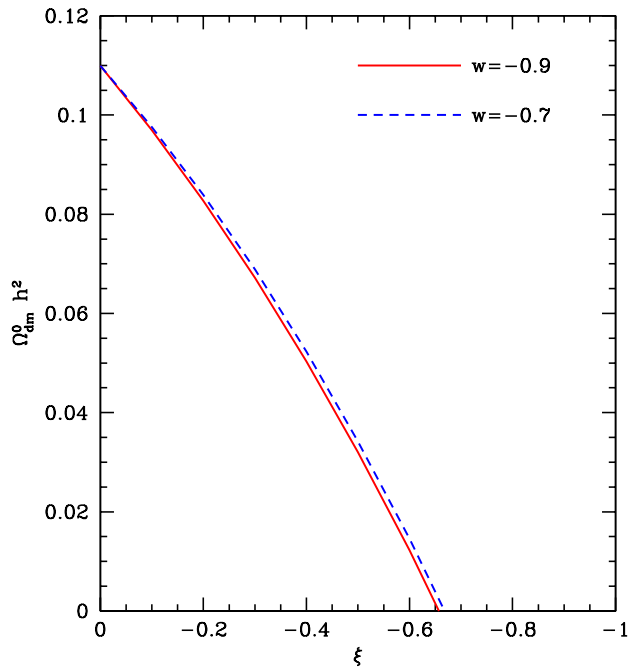}
  \caption{Present-day value of the dark matter energy density  of the universe for the class II DMvel and DEvel models characterized by $Q=\xi {\cal H} \rho_{\rm de}$ as a function of the parameter $\xi$, necessary to fit WMAP 5 year angular diameter distance data~\cite{Komatsu:2008hk} and the  physical dark matter and baryon densities at decoupling. The red solid (blue dashed) curve assumes an equation of state of the dark energy component $w=-0.9$ ($w=-0.7$).}
  \label{fig:beth}
\end{center} 
\end{figure}

In the class II DMvel and DEvel models we consider here, 
the energy-momentum transfer is completely negligible at the 
high redshift relevant to the CMB. 
Assuming a flat universe and perfect measurements of $\Omega_{\rm dm} h^2$,
$\Omega_{\rm b} h^2$, and the angular diameter distance to the last
scattering surface from CMB observations, the amplitude of $\xi$ is
degenerate with the physical energy density in dark matter today~\footnote{There is a corresponding degeneracy
  with $H_0$ that will be further quantified in
  \S \ref{subsec:shifts}.}. Figure~\ref{fig:beth} illustrates the values of $\Omega_{\rm dm}^{(0)}h^2$ necessary to fit WMAP 5 year angular diameter distance data, i.e. the first
 acoustic peak position~\cite{Komatsu:2008hk}, as a function of the dimensionless coupling constant $\xi$, taking into account that at the redshift of recombination $z_*$, $\Omega_{\rm
  dm}(z_*) \simeq (\Omega_{\rm dm}^{(0)} + \Omega_{\rm de}^{(0)} \xi/(3 w_{eff})) (1+z_*)^{3}$, or equivalently, that the dark matter do not evolve as $\Omega_{\rm dm}^{(0)} (1+z)^{3}$ in the coupled models under study. The values of  $\Omega_{\rm de}$ and $H$ are defined coherently with the flat universe assumption. The values of the
cosmological parameters $\Omega_{\rm dm}^{(0)}$ and $H_{0}$, obtained 
as a function of the coupling $\xi$, are 
presented in a Table in the Appendix~\ref{sec:cosm-param-accord}. 
The degeneracy direction depends only
slightly on the (constant) value of $w$ as shown in Fig.~\ref{fig:beth}. In the
following, we shall restrict ourselves to $\xi > -0.6$, to ensure
positive matter density at $z=0$. 

It may be puzzling that dark matter could become
negative, however, restricting ourselves to the $\xi$ range above,  
 this would happen in the future and by no means we can avoid it 
(the couplings $\xi < -0.6$ are already 
ruled out by a combined analysis of several cosmological
data\cite{Gavela:2009cy,Honorez:2009xt}).
An even more troubling feature is that in underdense regions $\rho
_{\rm dm}(x)$ may become negative well before $\Omega_{dm}$ does it
globally. This statement depends on the smoothing scale employed to
define $\rho_{\rm dm}(x)$. Therefore, it may happen that at some special 
point in the Universe, for any value of the coupling, at some point in
the past the current description makes the local dark matter density
negative. However, notice that this would happen when fluctuations become
highly non-linear, a regime which can only be explored with N-body
simulations. Rather than an inconsistency of the model this indicates that the
model should be seen as an  effective,  coarse grained description of
some more complex and fundamental model.

 At the level of perturbations, the phenomenology in these coupled models 
is simplified in the case where $c^2_{\rm de}$ is large, where dark energy
perturbations would be negligible. Moreover we will assume in this paper
that the perturbations in the expansion rate $\delta H$ can be
neglected, see Ref.~\cite{Gavela:2010tm} for more details. 
 In  this situation, the rate of energy transfer between dark matter and dark energy is nearly homogeneous. For an energy density transfer, for instance, 
$\alpha {\rho}_{\rm dm}$, the product ${\rho}_{\rm dm} \delta_{\rm dm}$ remains constant, while $\delta_{\rm dm}\rightarrow \delta_{\rm dm}/(1-\alpha)$.  
Therefore we expect very small changes to the lensing signal, 
$\propto \Omega_{\rm dm}\delta_{\rm dm}$. However, since $\delta_{\rm dm}$ grows, 
we do expect some enhancement to the growth of structure in these models. 
 
In a more quantitative approach, one can derive the growth equation for
dark matter in the Newtonian limit. We obtain:
\begin{eqnarray}
  \delta_{\rm dm}''&=&- B \frac{\delta_{\rm dm}'}{a}+ \frac32 \left(A \Omega_{\rm dm} 
  \frac{\delta_{\rm dm}}{a^2}+ \Omega_{\rm b}  \frac{\delta_{\rm b}}{a^2}\right)~,
\label{eq:grodm}
\end{eqnarray}
where
\begin{eqnarray}
B&=& 2-q+ (2-b)\xi \frac{\rho_{\rm de}}{\rho_{\rm dm}}~,\label{eq:B}\\
A&=& 1+\frac{2}{3}\frac{1}{\Omega_{\rm dm}}\frac{\rho_{\rm de}}{\rho_{\rm dm}}
\left[-\xi\left(1-q-3w\right)+\xi^2\left(\frac{\rho_{\rm de}}{\rho_{\rm dm}}+1\right)\right]~.
\label{eq:A}
\end{eqnarray}
Recall that $b=0$ for DEvel models and $b=1$ for DMvel models.
For negative coupling $\xi$, the Hubble friction term $B$ is suppressed and
the $A$ contribution to the source term in enhanced. This implies that
the dark matter growth in coupled models satisfying Eqs.~(\ref{eq:ourm}) will be larger than in uncoupled models. More generally, 
this feature is valid for any coupled model
in which $Q$ is directly proportional to the dark energy density and
$Q/\rho_{\rm de}$ is negative,  see {\it e.g.} Ref.~\cite{CalderaCabral:2009ja}. 

One could be tempted to interpret the change in the source term of
Eq.~(\ref{eq:A}) as the result of a fifth force for dark matter,  
the total attractive force between dark matter particles being driven by some
effective gravitational constant $G_{\rm eff}$ with $G_{\rm eff}/G= A$.
This interpretation would mean the violation of the weak equivalence
principle. However, such an interpretation is incorrect in the case of
$Q_\nu\propto u_\nu^{({\rm dm})}$, as the dark matter Euler
Eq.~(\ref{eq:thetames}) is not modified by the presence of the dark coupling and the velocities for dark matter and baryons are identical.

We analyze now  the linear growth in the three possible
cosmologies that lie along the one-dimensional degeneracy defined by
Fig.~\ref{fig:beth}. These three possible cosmologies, which have
identical background histories, are: {\it a)} an uncoupled, albeit
dynamical dark energy cosmology, with a varying equation of state $\tilde w(z)$
given by Eq.~(\ref{eq:wH}), with $H(z)$ given by Eq.~(\ref{eq:hz}) 
 and assuming that $\Omega_{\rm dm}$ in Eq.~ (\ref{eq:grodm}) scales with $a^{-3}$,
{\it b)} the coupled $Q\propto \rho_{\rm de}$ DMvel model with constant $w=-0.9$ given by the first equation of Eqs.~(\ref{eq:ourm}) and {\it c)} the 
coupled class II $Q\propto \rho_{\rm de}$ DEvel model with constant $w=-0.9$ given by the second expression of Eqs.~(\ref{eq:ourm}). While in the DMvel model
the baryon densities and velocities nearly trace those of the dark
matter, this is not the case for the DEvel model. This difference will be carefully explored in \S \ref{pecvel}-\ref{sec:wepvnew}. We consider observables
probed by lensing ($\propto\Omega_{\rm dm} \delta_{\rm dm}$, if one
ignores geometrical factors) and linear peculiar velocities ($\propto
\theta_{\rm dm}$)  see also \S~\ref{sec:obs}.

\begin{figure}
  \centering
 \includegraphics[scale=0.7, angle=0]{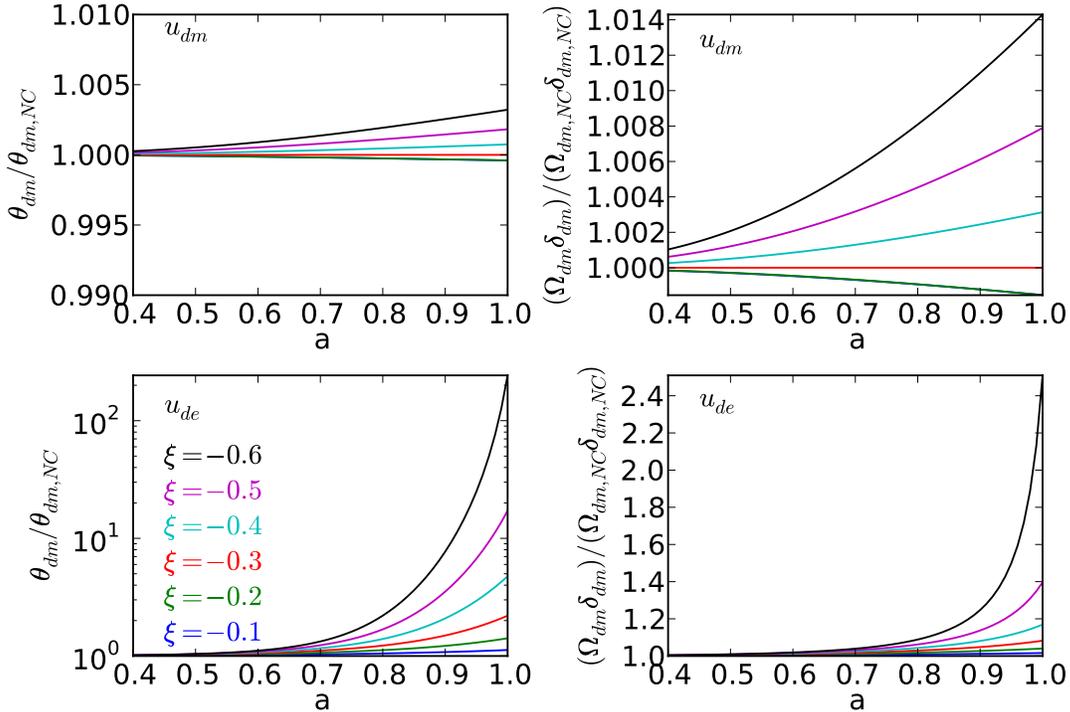}
  \caption{\label{fig:wztest}  The ratio of the amplitude of dark matter peculiar velocities, $\theta_{\rm dm}$, and the lensing signal, $\Omega_{\rm dm} \delta_{\rm dm}$ in coupled models compared to models with standard GR growth with identical expansion histories (labelled with NC for ``Not Coupled''). The curves show $\xi = 0, -0.1, ... -0.6$.  The upper left panel shows the velocity ratio in the DMvel model and the upper right panel shows the lensing ratio. The two bottom panels show the same but for the DEvel model.}
\end{figure}

Figure \ref{fig:wztest} shows the ratio of the values of $\Omega_{\rm dm}
\delta_{\rm dm}$ and $\theta_{\rm dm}$ in cases {\it b)} and {\it c)} 
to their values in an uncoupled
model with the same expansion history dictated by the relation in
 Fig.~\ref{fig:beth}, for six possible values of the coupling
$\xi$.  Let us emphasize that in Figure \ref{fig:wztest}, t
he quantity $\Omega_{\rm dm}\delta_{\rm dm}$ is derived using the solution 
for $\Omega_{\rm dm}$ of
Eq.~(\ref{eq:EOMm}) with non zero coupling, and $\delta_{\rm dm}$
satisfying the growth equation (\ref{eq:grodm}).  At the
background level, the cosmological parameters were chosen to match the
angular diameter distance to last scattering surface from CMB 
observations. $\Omega_{\rm dm, NC} \delta_{\rm dm,NC}$ refers to the
uncoupled model (Not Coupled) with the same expansion history than the coupled model. In the NC model $H(z)$ is constructed assuming that $\Omega_{\rm dm, NC}$ has the usual $a^{-3}$ dependence and the dark energy fluid is described by the effective equation of state of Eq.~(\ref{eq:wH}). The NC matter overdensity field $\delta_{\rm dm,NC}$ satisfies equation~(\ref{eq:eqf2}). An equivalent
approach has been followed to derive $\theta_{\rm dm}/\theta_{\rm dm,NC}$.
 
 For the DMvel  class II model (case {\it b)}), there are only tiny
changes to the velocity and lensing signals.
While there is a clear divergence of the
dark matter velocity and overdensity field in the DEvel class II model 
(case {\it c)}), we must be careful in considering how to apply 
observational constraints to
this model.  This enhancement of the velocity (for $\xi$ negative)
yields the fifth force effect which is only present in DEvel
 models (see also the $(1-b)$ term in Eq.~(\ref{eq:thetames})).
This effect will be described in more detail in \S \ref{sec:bulk}.  

In summary, in both the DMvel and the DEvel models the 
dark coupling can, in principle, be distinguished from a 
generic, uncoupled, dark energy described by $\tilde w(z)$ 
with perturbations growing as in standard GR.
Note however that in the DMvel model GR is not modified. Although extremely small in this example, this class of models  displays a mis-match between the reconstructed expansion history and the growth of structures, representing an exception to the commonly accepted interpretation that such a mis-match would be a tell-tale sign of modification of GR. In other words, if the  measured expansion history and growth  of structure are in agreement, this observation can be used to constrain deviations from GR. The conversely however is not necessarily true: a mis-match between expansion and growth could indicate deviations from GR or a dark coupling with GR unchanged.

\section{Low Redshift Observational Probes of Dark Coupling}
\label{sec:obs}

Coupled cosmologies, in order to fit the high-precision CMB data, predict values for the present-day cosmological parameters that differ substantially from the parameters values within non-interacting cosmologies. In other words, coupled cosmologies, in order to fit the high redshift (high $z$) observations, yield different expansion histories and different (shifted) values for  present-day  cosmological parameters. We show how to detect the induced shifts in the present-day value of some cosmological key parameters, as the Hubble constant $H_0$. 
Growth of structure probes are also explored and shown to be a
powerful tool: while current errors on local bulk flows are still
large, forecasted constraints on peculiar velocities offer a promising
avenue.
Let us mention that all the numerical results below were obtained
choosing the cosmological parameters (as, for instance, $\Omega_{(0)}{\rm dm} h^2$) accordingly to WMAP 5 year angular diameter distance
data~\cite{Komatsu:2008hk} as explained in section~\ref{lineargrowthblah} for the Class II models studied here. For Class I DMvel models these numbers are provided in Ref.~\cite{Valiviita:2009nu}. Also, changes in  e.g. last scattering
surface, matter-radiation equality, and growth history for
cosmological perturbations are fully taken into account by means of a modified version of CAMB~\cite{Lewis:1999bs} which includes coupling effect at the background and linear perturbation level.

\subsection{Background High-z vs Low-z  quantities mis-match: a worked example with  Hubble constant  measurements}
\label{subsec:shifts}

We first consider future CMB data (Planck) and future constraints on
$H_0$ (constraints from present cosmological data were already considered in
Ref.~\cite{Gavela:2009cy} using the cosmomc package~\cite{Lewis:2002ah}). As it is often the case for forecasting the errors on parameters achievable with future data we make use of the Fisher matrix approach. 

Let us focus on the DMvel class II dark coupled models given by   $Q=\xi {\cal H} \rho_{\rm de}$ (i.e. the first
equation of Eqs.~(\ref{eq:ourm})) and 
calculate the shifts induced by the (neglected) coupling $\xi$ on the present-day value 
of the different cosmological parameters.
We apply the technique of Ref.~\cite{Heavens:2007ka} to compute the
expected shifts at zero redshift on the cosmological parameters due to
the presence of non-zero coupling in the data and of setting coupling
to zero in the analysis. The authors of \cite{Heavens:2007ka}
developed a method  which exploits the Fisher information matrix
formalism. We briefly summarize their method in the context of
interacting models. Let us assume that we have two competing models:
the $M^\prime$ model (a $\Lambda$CDM universe) versus the M model (the
DMvel coupled cosmology considered here). The method assumes that the
$n^\prime$ parameters of model $M^\prime$ are common to $M$, which has
$p=n-n^\prime$ extra parameters in it. These extra parameters are
fixed to fiducial values in $M^\prime$. For the form of dark coupling
we illustrate here (the application to other coupled models is
straightforward), $p=1$, being the extra parameter the dimensionless
coupling $\xi$ of the DMvel interacting model, which is set to $\xi=0$
in the $M^\prime$ model ($\Lambda$CDM). If the true underlying model
turns out to be the $M$ model (coupled cosmology), in the $M^\prime$
model, the maximum of the expected likelihood will not be at the
correct values of the parameters. The $n^\prime$ parameters shift from
their true values to compensate the fact that, effectively, the
coupling $\xi$ additional parameter is being kept at an incorrect
fiducial value (i.e. is set to zero). If this (incorrect) fiducial
value differs by $\delta \phi_\xi$ from its true value, the others are
shifted by an amount~\cite{Heavens:2007ka}       
   \begin{equation}
\delta \theta_\alpha=-(F^{\prime -1})_{\alpha \beta}G_{\beta \xi} \delta \phi_\xi~,
\end{equation}
where $\alpha,\beta=1...n^\prime$, $F^\prime$ is the Fisher matrix for
the model $M^\prime$ ($\Lambda$CDM) and $G_{\beta \xi}$ is a submatrix
of the $M$ model (interacting cosmology) Fisher matrix. The set of
$n^\prime$ parameters we use here to describe the $M^\prime$ model are
the current baryon and dark matter energy densities $\Omega_{\rm b}
h^2$ and $\Omega_{\rm dm} h^2$, the current value of the Hubble
parameter $h$, the amplitude $A_s$ of the primordial scalar spectrum,
the scalar spectral index $n_s$ and the equation of state of the dark
energy component $w$. 
We have computed the shift in the $n^\prime$ parameters for a CMB
experiment with the specifications of
Planck~\footnote{www.rssd.esa.int/PLANCK}. The CMB Fisher matrix has
been computed following Ref.~\cite{Verde:2005ff} using a modified
version of CAMB which includes the effect of the coupling at the
background and linear order perturbation level. The Fisher matrix in the 
 $M^\prime$ cosmology contains six parameters: $\Omega_{\rm dm}h^2$,$\Omega_{b}h^2$, $H_0$, $n_{\rm s}$, $w$ and $A_{\rm s}$, being $A_{\rm s}$ is the amplitude of the 
primordial spectrum and $n_{\rm s}$ the spectral index. The Fisher matrix in the 
coupled cosmology will contain an extra parameter, the dimensionless coupling 
$\xi$.

If we were living in a coupled Universe and future CMB Planck data were
to be (wrongly) fitted to a $\Lambda$CDM cosmology, the value of $H_0$
would shift by $d\ln H_0/d\xi=0.3$. This is the result of our Fisher
analysis. However if the CMB data were to be
fitted to a model where $w$ is also a free parameter, new degeneracies
open up and  $d\ln H_0/d\xi$ increases.  Thus if the Hubble
constant $H_0$ is determined independently   with an uncertainty below
$3\%$,  small values of coupling could be ruled out.
A number of experiments (HST, Spitzer, GAIA and
JWST~\cite{Freedman:2010xv}) are expected to measure $H_0$ with $2\%$
uncertainty in the next decade. Consequently,  coupled models could be
highly disfavoured over the next decade by the combination of CMB
Planck data with precise measurements of the Hubble constant $H_0$.
The other cosmological parameters will also be shifted from their true
values but by amounts expected to be smaller  than their  combined
statistical and systematic errors.

The shifts for $H(z)$ (at $z>0$), as one may expect, are maximal at
$z=0$, thus forthcoming BAO observations  which constrain $H(z)$ with
$\sim \%$ precision but only at a $z$ where the volume element per
unit solid angle is near maximal, will not offer significant  further
improvement. 

To conclude, in this section we have provided quantitative forecasts of how
future constraints on $H_0$ can help in constraining the coupling between
the dark sectors.

\subsection{Effects on the skewness}
\label{sec:skewness}
It is well known that the skewness of the density field is relatively
insensitive to the expansion history (and thus the cosmological
background) but is very sensitive to the growth of structure: mildly
non-linear gravity  leaves a distinct signature on the three-point
correlation function (and therefore on the skewness) and galaxy bias
also alters the skewness of the  galaxy density field compared with
the skewness predicted for the dark matter distribution. Here we
consider the skewness as a test for  the effects of  dark  coupling,
following Refs.~\cite{Amendola:2003wa,Amendola:2004wa}. A large effect
on the skewness is a clear indication  that the growth of
perturbations is modified with respect to the standard
expectation. Skewness tests are  also affected by the ambiguity of
whether galaxies trace the baryon or the dark matter
distribution.

Skewness is defined as the third order moment of the galaxy distribution
\begin{equation} S_{3}=\frac{<\delta (x)^{3}>}{(<\delta (x)^{2}>)^{2}}\, ,\label{eq:s3}\end{equation}
where $\delta (x)$ is the density contrast at the point $x$.
 In an Einstein de Sitter cosmology the  skewness  is predicted to be $34/7$. This value
has been shown to be independent of the nature of the dark energy component~\cite{Kamionkowski:1998fv} or the background cosmology~\cite{bouchet,catelan} and it is only very mildly dependent on the shape of the primordial power spectrum (given current constraints on the spectral index). Here, we study possible deviations of the skewness from its standard value in the context of coupled cosmologies. Following Ref.~\cite{Kamionkowski:1998fv} we compute the perturbation evolution in second order perturbation theory in coupled cosmologies, and compute the skewness parameter as 
\begin{equation}
S_{3}=4+2\epsilon=\frac{34}{7}+\frac{6}{7}\left(\frac{7}{3}\epsilon-1\right)~,
\end{equation}
where $\epsilon$ is a function of the first and second-order time-dependent components
of density contrast $\delta_{(1)}(t)$ and $\delta_{(2)}(t)$, and reads
\begin{equation}
\epsilon=\frac{\delta_{(2)}(t)}{\delta^2_{(1)}(t)}~.
\end{equation}
 $\delta_{(1)}(t)$ is the solution of the  growth equation at first
order in perturbation  which corresponds to Eq.~(\ref{eq:eqf2}) for
uncoupled or DMvel class I models and to Eq.~(\ref{eq:grodm}) for
DMvel $\&$ DEvel class  II considered here. Note, as we illustrate in the
following, that the evolution equation
for the second order density perturbation  $\delta_{(2)}(t)$  for the matter fluids $\alpha=\rm{dm, b}$ depends as well on
$\delta_{(1)}(t)$.

 Let us first analyze the
case of  $Q\propto \rho_{\rm dm}$ (class I) models.  Ref.~\cite{Amendola:2004wa}
has shown that the skewness deviations in DEvel class I cosmologies
are $\sim 1\%$ for values of the coupling parameter satisfying current CMB
constraints. For uncoupled and DMvel class I interacting models, the equation for the second order density perturbation evolution for the
matter fluids $\alpha=\rm{dm, b}$ reads 
  \begin{equation}
\delta_{(2),\rm dm}''= -(2-q)\frac{\delta_{(2),\rm dm}'}{a} + \frac{3}{2}\left[\Omega_{\rm dm}\left(\frac{\delta_{(2),\rm dm}}{a^2}+ \frac{{\delta}^2_{(1),\rm dm}}{a^2}\right) + \Omega_{\rm b}\left(\frac{{\delta}_{(2),\rm b}}{a^2}+ \frac{{\delta}_{(1),\rm dm}{\delta}_{(1),\rm b}}{a^2}\right)\right]~.
\end{equation}
For the DMvel class I interacting model of Eq.~(\ref{eq:maart}), the
deviations of the skewness for both the cold dark matter and baryon
distributions from its standard value is smaller than $1\%$ for
the WMAP best fit of Ref.~\cite{Valiviita:2009nu} and
thus hard to be measured, even with upcoming future galaxy
surveys. Recall that for DMvel class I models the continuity
and Euler equations are exactly the same as for non interacting cases (after
neglecting dark energy perturbations), and the only change in
the growth is only introduced via the modified background
evolution of ${\mathcal H}$ and $\rho_{\rm
  dm}$~\cite{CalderaCabral:2009ja,Koyama:2009gd}.

For the DMvel and DEvel coupled class II models with $Q= \xi {\mathcal
  H} \rho_{\rm de}$ of  Eqs.~(\ref{eq:ourm}), the second order
growth equation for the dark matter fluid in the Newtonian limit is
given by: 
\begin{equation}
\delta_{(2),\rm dm}''= -B \frac{\delta_{(2),\rm dm}'}{a} + \frac{3}{2}\left[\Omega_{\rm dm}\left(\frac{ A \delta_{(2),\rm dm}}{a^2}+ \frac{{\delta}^2_{(1),\rm dm}}{a^2}\right) + \Omega_{\rm b}\left(\frac{{\delta}_{(2),\rm b}}{a^2}+ \frac{{\delta}_{(1),\rm dm}{\delta}_{(1),\rm b}}{a^2}\right)\right]~,
\label{eq:eqf2de} 
\end{equation}
where $A$ and $B$ are given by Eqs.~(\ref{eq:A}) and (\ref{eq:B})
respectively. Let us recall that in order to obtain the growth
equations for  $Q= \xi {\mathcal H} \rho_{\rm de}$ models we have
neglected the contributions of $\delta_Q$.
The additional terms proportional to $\xi \delta$, $\xi \delta'$ in
Eqs.~(\ref{eq:grodm})  and (\ref{eq:eqf2de}) yield large effects on the
cold dark matter distribution, and we observe $\sim 10\%$ deviations
in the dark matter skewness. For the baryon distribution, the effects
are below $1\%$.

 As already pointed out in
\S~\ref{lineargrowthblah}, for these two DMvel and DEvel class II
models, the cold dark matter perturbation evolution can differ
significantly from the baryon evolution, i.e. the changes in the
growth of perturbations at linear and second order in these two models are not
just due to a different background evolution. Even if the skewness
effects in the cold dark matter distribution for these interacting
models are large, it is not clear if they can be  measured by galaxy
surveys. Remember that in $\Lambda$CDM baryons and dark matter
interact gravitationally in the same way at large scales so that the
problem does not appear.

If observations of the skewness of the galaxy distribution probe the
baryon distribution, then skewness tests are not useful in
constraining the dark coupling. Indeed, the presence 
of the coupling term only give rise to a $1\%$ effect which is small 
compared with any foreseeable error-bar.
If skewness observations (maybe via weak lensing, although recall that
weak lensing  probes the combination $\delta_{\rm dm} \Omega_{\rm dm}$ and not $\delta_{\rm dm}$ or $\Omega_{\rm dm}$ alone) can yield the dark matter skewness, then 
this approach may be useful given the expected $10\%$ effect.


\subsection{Peculiar Velocities}
\label{pecvel}
The velocity field offers a test of the growth of structure, which is
complementary to galaxy clustering, as, for example, is less sensitive
to non-linearities and bias. Below we consider constraints that can be
obtained from present bulk flow data in the local universe and
forecasted constraints achievable from future galaxy redshift
surveys. These forecasts should be considered complementary to those
presented in
Refs.~\cite{Amendola:2006dg,Wang:2006qw,Guo:2007zk,Olivares:2007rt,Feng:2008fx,He:2008tn,He:2009pd,Gavela:2009cy,Valiviita:2009nu,Gavela:2010tm}.
Throughout this section we will fix the CMB observables angular diameter distance to the last scattering surface   and the physical  dark matter and baryon density at decoupling.

\subsubsection{Constraints from peculiar velocities: Local bulk flows}
\label{sec:bulk}
Recently Watkins {\it et al.}~\cite{Watkins:2008hf} have reported the
observation of anomalously large bulk flows on 100 $ h^{-1}$ Mpc
scales. In a Gaussian window of radius $50 h^{-1}$ Mpc, they find a coherent 
bulk motion of $407\pm 81$~km/s in conflict with the $\Lambda$CDM expectation
of $\sim 200$~km/s at the $2\sigma$ level. Reference~\cite{Ayaita:2009qz} pointed out that these
results, if confirmed, would favour models with a growth of
perturbations larger than $\Lambda$CDM model predictions. 

We first clarify the relation between the density and velocity fields in a general context where coupling is allowed.  We again adopt a CMB-centric view: the amplitude of fluctuations at the epoch of recombination are well-measured, and so in the analysis that follows, we consider those to be fixed.  We denote the amplitude of the density perturbation $A_{\bf k}$ for a particular ${\bf k}$-mode.  The solution of the linear perturbations evolution (Eqs.~(\ref{eq:deltab}) - (\ref{eq:thetaees})),  which we refer to by ${\tilde \theta}_{\rm dm/b}(z)$ and ${\tilde \delta}_{\rm dm/b}(z)$, describe the redshift evolution of the amplitude of each independent ${\bf k}$ mode:
\begin{eqnarray}
\delta_{{\bf k}, {\rm dm/b}}(z) & = & A_{\bf k} \frac{{\tilde \delta_{\rm dm/b}}(z)}{{\tilde \delta_{\rm dm/b}}(z_{\rm CMB})} \\
\theta_{{\bf k}, {\rm dm/b}}(z) & = & A_{\bf k} \frac{{\tilde \theta_{\rm dm/b}}(z)}{{\tilde \theta_{\rm dm/b}}(z_{\rm CMB})}\,,
\end{eqnarray}
where we have emphasized that when coupling is present, the growth of baryon and matter density and velocity perturbations are not necessarily the same.  In the uncoupled case in the Newtonian limit ($k\gg {\cal H}$) and assuming that $\hat c_{s\,{\rm de}}^2=1$, baryon and matter perturbations trace each other, and the velocity and density perturbations are related by:
\begin{equation}
\label{u_time}
{\theta}_{{\bf k},{\rm dm/b}} = -aH\frac{A_{\bf k}}{{\tilde \delta_{\rm dm/b}}(z_{\rm CMB})} \frac{\partial {\tilde \delta}_{\rm dm/b}}{\partial \ln a} \; ({\rm uncoupled})\,.
\end{equation}
In general, the relation between the growth equation solutions ${\tilde \delta}_{dm,b}$ and ${\tilde \theta}_{dm,b}$ is not so simple.  Nevertheless, one can eliminate the original mode amplitude $A_{\bf k}$ by defining

\begin{eqnarray}
\label{tildefdef}
{\tilde f}_{\rm dm,b}(z) \equiv -\frac{{\tilde \theta}_{\rm dm,b}(z)/{\tilde \theta}_{\rm dm,b}(z_{\rm CMB})}{{aH\tilde \delta}_{\rm dm,b}(z)/{\tilde \delta}_{\rm dm,b}(z_{\rm CMB})}.
\end{eqnarray}

In the non-coupled case, ${\tilde f} = \partial \ln {\tilde \delta}/\partial \ln a$ (the same for baryons and dark matter), while in class II models ($Q \propto \rho_{\rm de}$) for both DEvel and DMvel, ${\tilde f}_{\rm dm} = \partial \ln {\tilde \delta_{\rm dm}}/\partial \ln a + \xi \rho_{\rm de}/\rho_{\rm dm}$.  Finally, we can write the linear velocity field in terms of the density field as 
\begin{equation} 
{\vec v}_{\rm dm,b}(k,a)=\frac{i {\tilde f}_{\rm dm,b} a H
\delta_{\rm dm,b}(k,a){\vec k}}{k^2}.
\label{eq:vdelta}
\end{equation}
Following Ref.~\cite{Ayaita:2009qz}, the expected mean square velocity is
\begin{eqnarray}
    \left< u^2 \right>&=& \frac{1}{2\pi^2} \int_{0}^{\infty} {\rm d} k\, k^2 P_v(k)
	|\tilde W(k)|^2 \label{u_exact}
    \label{eq:Pv_def} \\
    & = & \frac{{\tilde \theta(z)}^2 }{2\pi^2{{\tilde \theta}^2(z_{\rm CMB})}} \int_{0}^{\infty} {\rm d} k\, \left<A_{\bf k} A^{\star}_{\bf k}\right> 
	|\tilde W(k)|^2\,,
  \end{eqnarray}
  where $P_v(k)$ is the peculiar velocity power spectrum and $\tilde W(k)$ is
the  Fourier transformed spherically symmetric window function~\footnote{For our numerical calculations, we use a Gaussian window function $\tilde W(k)= e^{- k^2 R^2/2}$ of radius $R=50 h^{-1}$~Mpc.}. 
In the second equality, the term outside the integral depends only on the normalized growth of velocity perturbations, ${\tilde \theta}(z)/{{\tilde \theta}(z_{\rm CMB})}$, while the value of the integral depends on the power spectrum of initial fluctuations.\footnote{This is not precisely true since $\left<A_{\bf k} A^{\star}_{\bf k}\right>$ is constrained tightly by the CMB in physical units ($k$ in Mpc$^{-1}$), while the radius of the window function is fixed at $50 h^{-1}$ Mpc.  We account for this dependence on $h$ in our calculations.}  This formula shows that bulk flows would be larger if the growth of perturbations is enhanced by some mechanism. In coupled dark matter-dark energy cosmologies, the growth factor can be modified in several ways~\cite{Koyama:2009gd} and therefore the predictions for the bulk flow may differ from those obtained in the context of a $\Lambda$CDM universe. Here we explore if coupled models could be favoured by bulk flow measurements.

For the DMvel $Q\propto \rho_{\rm dm}$ interacting model of Eq.~(\ref{eq:maart}) both
the continuity and Euler equations are unchanged and therefore, the
expression for the uncoupled velocity amplitude 
$f=\partial \ln{\tilde \delta}/\partial \ln a$ will still hold. 

Ref.~\cite{Valiviita:2009nu} showed that in this model, CMB
data alone disfavor positive values of $\Gamma$  but still allow large
negative values. For their CMB best fit model with $\Gamma/H_0=-0.3$, we
obtain a bulk flow of $171$~km/s. In principle, positive, higher
values of $\Gamma$ will predict larger values for the bulk
flows. However, these values of $\Gamma$ would not provide a good fit
to CMB and/or BAO and SN data, see the analysis presented in
Ref.~\cite{Valiviita:2009nu}. Therefore, bulk flow observations do not
indicate a preference for this particular coupled model.
\begin{figure}[h!]
\begin{center}
\includegraphics[height=.4\textheight]{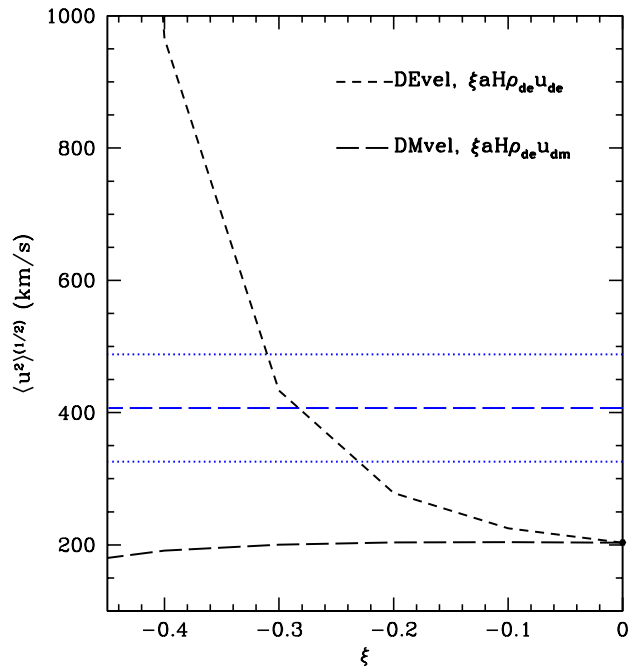}
  \caption{The short (long) dashed curve shows the bulk flow predictions for the DEvel (DMvel) class II interacting model with $Q=\xi {\cal H}\rho_{\rm de}$. The cosmological parameters have been chosen to fit WMAP 5 year angular diameter distance data~\cite{Komatsu:2008hk}. The dashed (dotted) blue lines represent the measured value of $407\pm 81$~km/s, and the circle depicts the $\Lambda$CDM model prediction for the bulk flows $\langle u^2\rangle^{1/2}= 203$~km/s.} 
   \label{fig:bulk}
\end{center} 
\end{figure}

Figure \ref{fig:bulk} shows the bulk flow results for the DMvel and
DEvel class II models ($Q \propto \rho_{\rm de}$) as a function of the coupling $\xi$. The concordance  $\Lambda$CDM cosmology predicts a bulk flow of $\langle u^2\rangle^{1/2}= 203$~km/s. As we have
seen in \S \ref{lineargrowthblah}, the DMvel model has dark matter
peculiar velocities which are similar to those of non interacting
models, since the Euler equation is unmodified. This is not the case
for the DEvel model, in which the Euler equation changes, as can be
 seen from Fig.~\ref{fig:bulk}: values of the coupling $\xi< -0.35$ 
lead to an effective 3 $\sigma$ level deviation from observations.

However, this model must be handled with care since the dark matter and baryon perturbations are not equal.  This large bias between the dark matter, baryon peculiar velocities and overdensities was shown in \S
\ref{lineargrowthblah}.  Though smaller in amplitude, this feature is
also seen in the coupled quintessence model explored with $N$-body
simulations in Ref. \cite{Maccio:2003yk,Baldi}. While it is at first
unclear which field the galaxies trace, we argue here  that for $ \xi  > -   1$ the velocities probed by,
e.g., luminous red galaxies should be those of the {\em dark matter},
at least in the class II model considered in \S
\ref{lineargrowthblah}.  {\it It is generally accepted that the massive} luminous red galaxies (LRGs) were in place by
redshift $z>1$ (see Refs.\cite{dunlop, Spinradetal97,cowie,heavens, thomas, panter, jimenez, treu} and references therein).  
 At this redshift, the energy-momentum exchange is negligible.  
Therefore, galaxy formation should proceed as expected in $\Lambda$CDM -- in the
 potential wells of dark matter halos.  Moreover, we know that luminous red galaxies
 occupy massive dark matter halos today from weak lensing measurements
 \cite{mandelbaum:2006}.  
 
 To demonstrate that these galaxies remain bound in the potential wells of their dark matter halos, we need to demonstrate that the differential acceleration of the dark matter and baryons due to the WEPV ``fifth force'' in this model is much smaller than the force binding the galaxy to the dark matter halo.  As discussed in \S \ref{sec:4moment-trsf}, the choice of DEvel and $Q \propto \rho_{de}$ implies a fractional increase in the dark matter peculiar velocity equal and opposite to the fractional change in energy density as energy is transferred from dark matter to dark energy.  This corresponds to an acceleration on a dark matter halo with peculiar velocity $v$ as
\begin{equation}
\left(\frac{dv}{dt}\right)_{\rm WEPV} = v \times \frac{1}{m} \frac{dm}{dt}.
\end{equation}
A dark matter halo has $\rho_{\rm halo} \sim 200 \rho_{\rm crit}$, so even for large couplings that correspond to depleting all of the mean dark matter density $\bar{\rho}_{\rm dm}$ in the universe in the past 1 Gyr, $1/m \; dm/dt < 1/200 \; {\rm Gyr}^{-1}$.  We take $v$ to be the expected rms linear theory bulk flow, $\sim 500$ km/s, and find $\left(\frac{dv}{dt}\right)_{\rm WEPV} < 8 \times 10^{-14}$ m s$^{-2}$.  For a normal Newtonian orbit, the gravitational acceleration is
\begin{equation}
\left(\frac{dv}{dt}\right)_{\rm gravity} = \frac{GM}{r_{\rm eff}^2} = \frac{GM}{r_{\rm vir}^3} \frac{r_{\rm vir}}{\kappa^2} \sim \frac{100 H_0^2 r_{\rm vir}}{\kappa^2}\,,
\end{equation}
where $M$ is the mass of the halo and $r_{\rm eff}$ is some fraction $\kappa < 1$ of the virial radius $r_{\rm vir}$ of the halo.  For $r_{\rm vir} \sim 0.5$ Mpc appropriate for these galaxies, we get $\left(\frac{dv}{dt}\right)_{\rm gravity} > 100 \left(\frac{dv}{dt}\right)_{\rm WEPV}$.  We conclude that under these assumptions, the
 galaxy peculiar velocities will trace the dark matter peculiar
 velocity field, and galaxy peculiar velocity observational results
 can be generally applied to interacting (DMvel or DEvel) class I and class II 
models.

 \subsubsection{Constraints from  peculiar velocities: redshift space distortions}
\label{sec:Kaiser}
Even in the linear regime, peculiar velocities make the  linear galaxy redshift-space power spectrum  $P_{\rm s} ({\bf k})$ anisotropic when the underlying real space linear power spectrum $P_{\rm dm}(k)$ is isotropic. These anisotropies go under the name of redshift-space distortions.
In linear theory and under  the flat-sky, distant observer approximation, we can generalize the results of \cite{Kaiser:1987qv,songpercival} to coupled models by simply replacing $f \equiv \partial \ln {\tilde \delta}/\partial \ln a$ with ${\tilde f}$ defined in Eq.~(\ref{tildefdef}):
\begin{equation}
  \delta^{\rm gal}_{\rm s}({\bf k},z)=(b_{\rm gal}+{\tilde f} \mu^2) \delta_{\rm dm}(k,z)~,\label{psm}
\end{equation}
where $\mu$ is the cosine of the angle between the wave vector ${\bf k}$ and the line of sight of a distant observer, $b_{\rm gal}$ the bias relating galaxy with dark matter overdensities in real space, i.e. $b_{\rm gal} \equiv {\delta_{\rm g}}/{\delta_{\rm dm}}$, and $\delta_{\rm dm}(k,z)$ is the real space (isotropic) linear dark matter overdensity.  Note that we assume in this section as well as in the previous one that galaxies trace the {\em dark matter} velocity fields rather than that of the baryons, since galaxies reside in dark matter halos.  Therefore, in the notation of the previous section, galaxy redshift surveys constrain ${\tilde f} \delta_{\rm dm} =({\tilde \theta}_{\rm dm}(z)/{\tilde \theta}_{\rm dm}(z_{CMB})aH) \left<A_{\bf k} A^{\star}_{\bf k}\right>^{1/2}$ using the $\mu$ dependence in Eq. (\ref{psm}).  Note that in the literature (e.g., \cite{songpercival}) the redshift space distortions constraint is often written instead as $f\sigma_8$, even though this probe is only sensitive to the amplitude of the velocity field, and not the amplitude of density perturbations directly.  We follow this labelling convention, but replacing $f \rightarrow {\tilde f}$ to account for the effects of dark coupling.
Since coupled cosmologies can modify the growth of density and velocity perturbations in 
a significant way, 
one can use the measurements of redshift space distortions to constrain the
strength of the dark sector interaction.

We first consider the DMvel class I coupled model of Eq.~(\ref{eq:maart}). For
this form of coupling, as previously stated, both the continuity and
Euler equations remain unchanged and in this case ${\tilde f} = \partial \ln \delta/\partial \ln a$.  The only imprint of
the dark sector interaction in the linear growth of perturbations for
this first interacting model will therefore arise from the background
quantities ${\mathcal H}$ and $\rho_{\rm  dm}$~\cite{CalderaCabral:2009ja,Koyama:2009gd}.  

We also consider here the DMvel and DEvel class II interacting
cosmologies of Eqs.~(\ref{eq:ourm}). For these two cosmologies, the
linear growth of the matter and velocity overdensities have extra contributions  
and therefore, the changes in the redshift space distortions observable will not arise exclusively from background
quantities. 

The results of \S~\ref{lineargrowthblah} indicate that  the impact of coupling on velocity perturbations will predominantly be through the change in background in DMvel models,
while sufficiently large couplings cause exponential growth of velocity perturbations in the DEvel model.

\begin{figure}[h!]
\begin{center}
\hspace*{-1.cm}  
\begin{tabular}{cccc}
\includegraphics[height=.25\textheight]{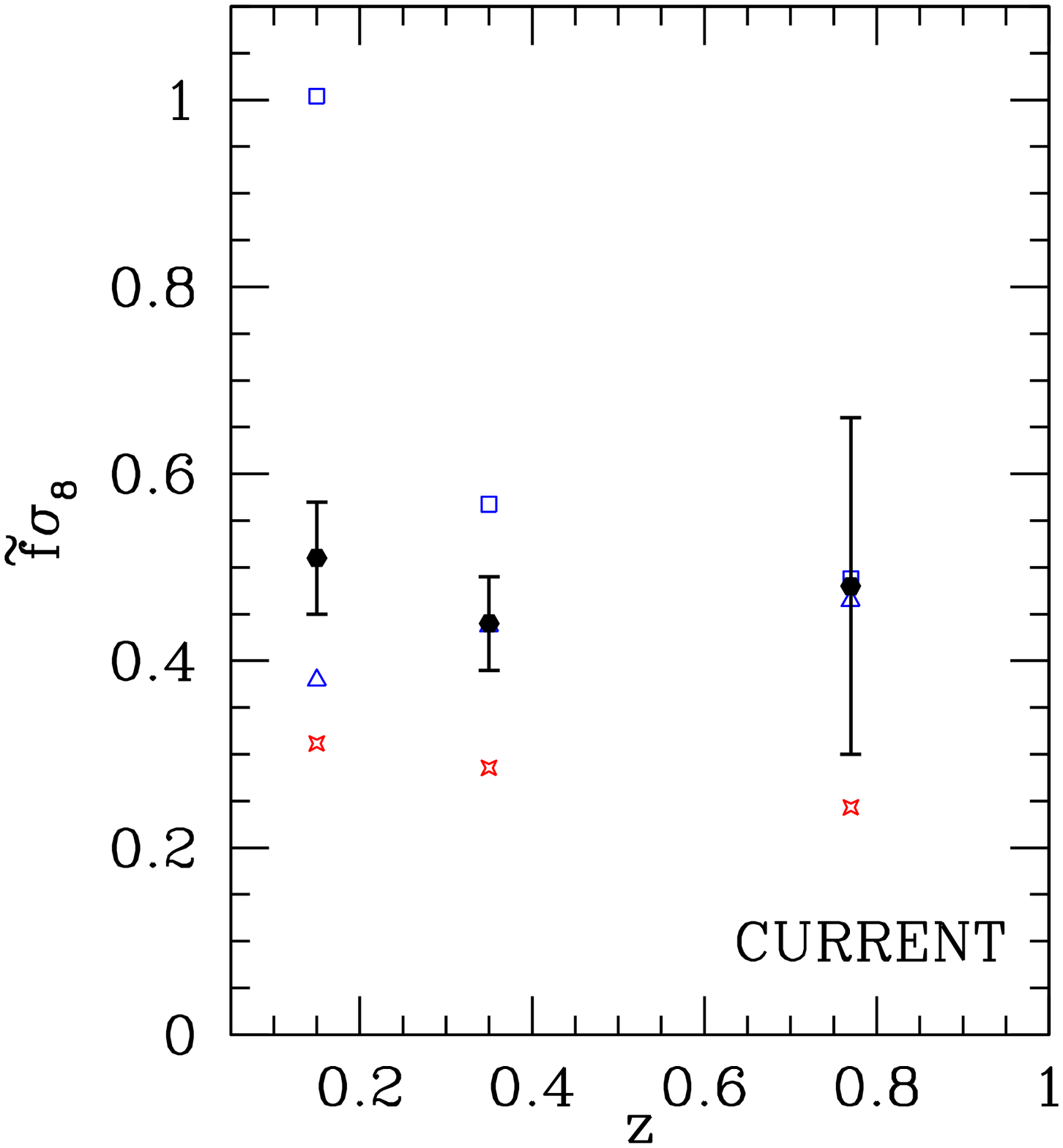}&
\hspace*{-.7cm}\includegraphics[height=.25\textheight]{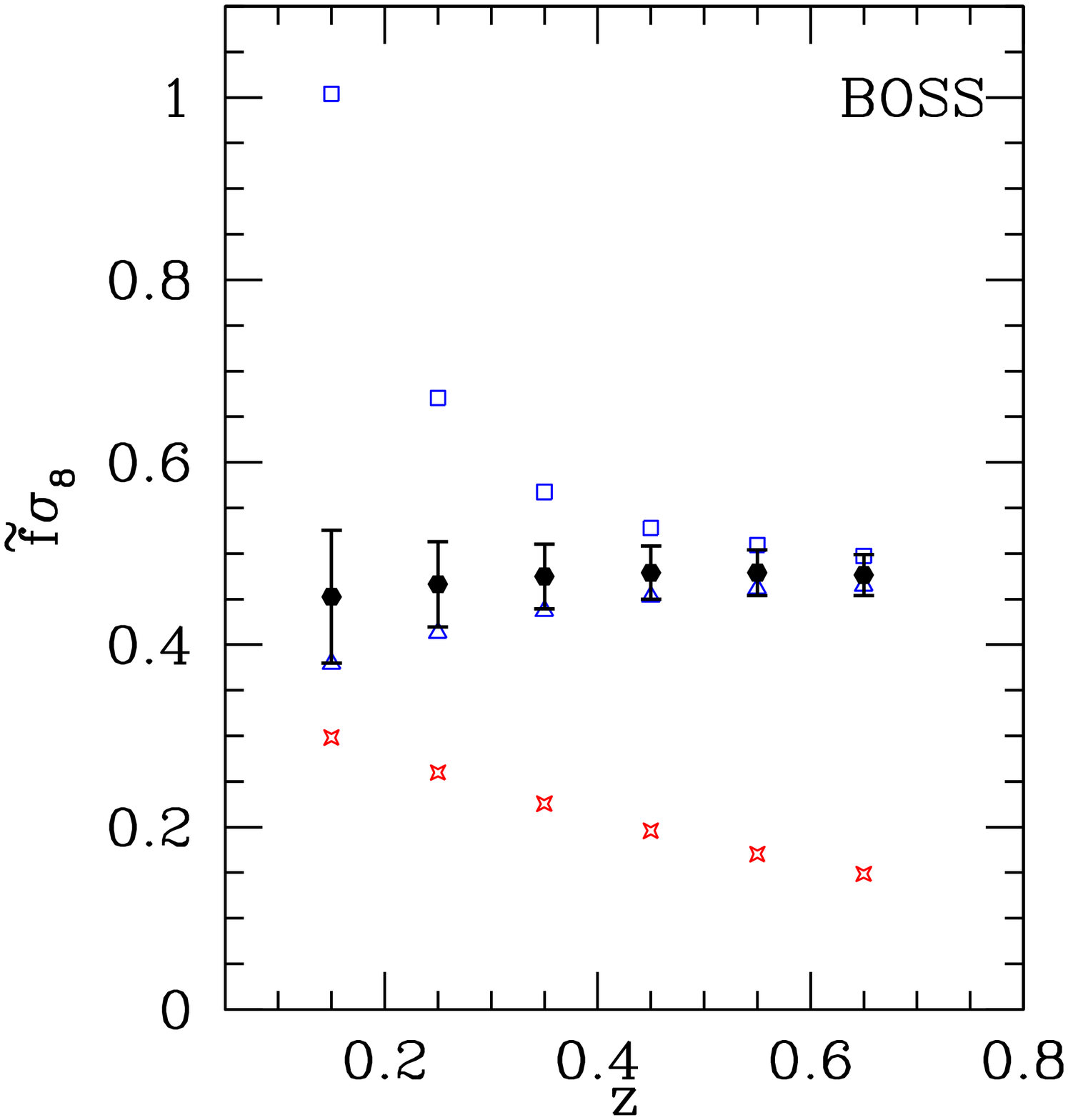}&
\hspace*{-.7cm}\includegraphics[height=.25\textheight]{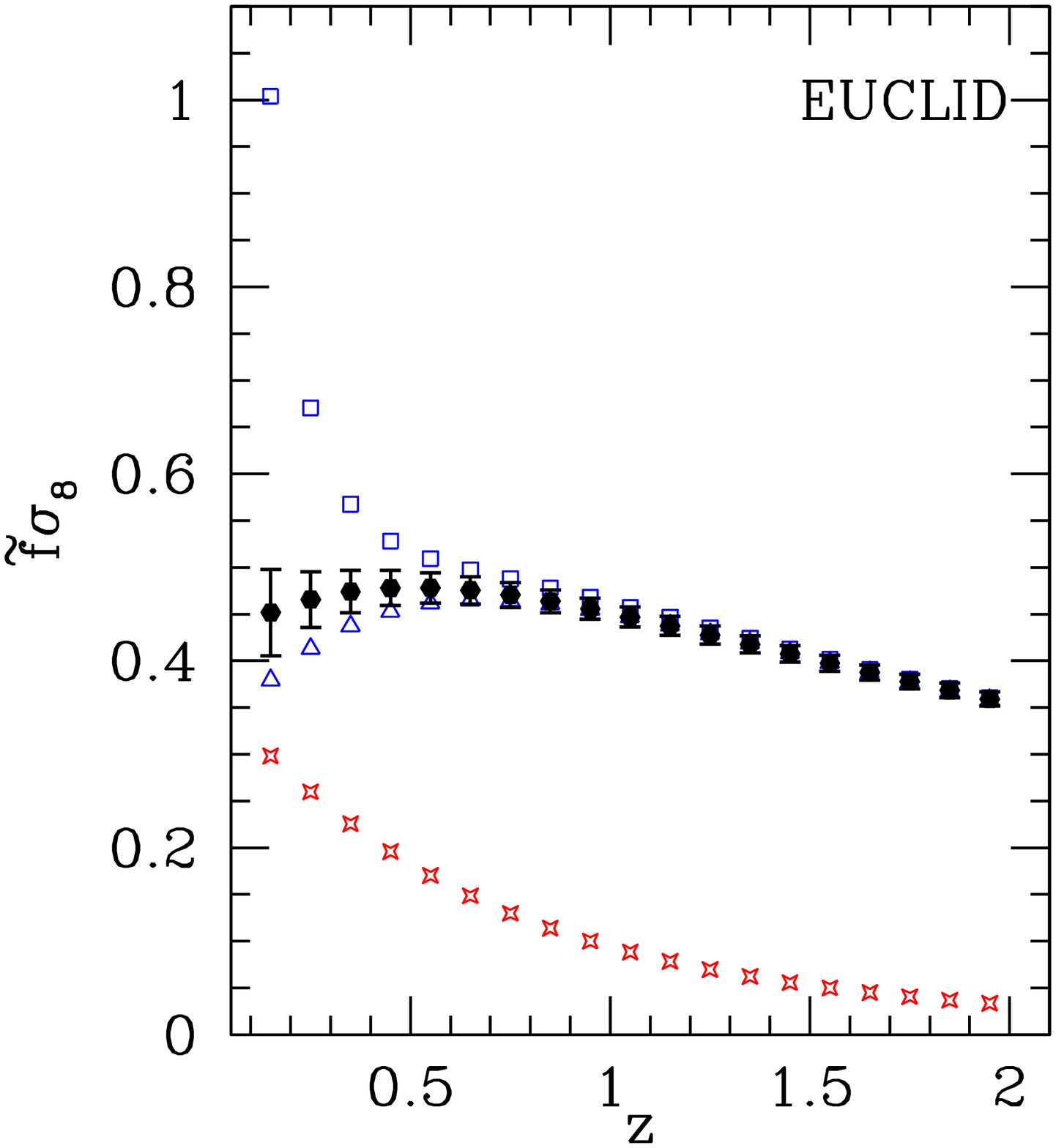}
 \end{tabular}
  \caption{The red crosses denote the ${\tilde f}\sigma_8$ values for the model of Eq.~(\ref{eq:maart}), for $\Gamma/H_0=-0.3$ and cosmological parameters fixed to model 2 of Ref.~\cite{Valiviita:2009nu}. The blue triangles (squares) depict the ${\tilde f}\sigma_8$ expectations for a DMvel (DEvel) class II model with $Q=\xi {\cal H} \rho_{\rm de}$ for $\xi=-0.5$. The points in the left panel show current measurements of ${\tilde f}\sigma_8$. The middle and right panels show the expected ${\tilde f}\sigma_8$ errors for BOSS and EUCLID-like surveys, respectively, assuming a fiducial $\Lambda$CDM cosmology.}
  \label{fig:fs8}
\end{center} 
\end{figure}
 Figure \ref{fig:fs8} illustrates the current and future constraining power of this method, where the red crosses show the ${\tilde f}\sigma_8$ predictions for the DMvel class I coupled model of Eq.~(\ref{eq:maart}), for $\Gamma/H_0=-0.3$, where the cosmological parameters have been fixed to their WMAP 5 year data best fit values (see model 2 of Ref.~\cite{Valiviita:2009nu}). The blue triangles (squares) refer to the predictions of the DMvel (DEvel) class II model explored here, with $\xi=-0.5$ and $w=-0.9$ (the remaining cosmological parameters are chosen accordingly to WMAP 5 year angular diameter distance data~\cite{Komatsu:2008hk}). The black circles in the first panel of Fig.~\ref{fig:fs8} represent current ${\tilde f}\sigma_8$ measurements~\cite{Song:2008qt}, in the other two panels they show the ${\tilde f} \sigma_8$ forecasts for BOSS and EUCLID-like galaxy surveys, assuming a $\Lambda$CDM fiducial cosmology. In order to forecast the errors from these two BAO galaxy surveys, we have used the Fisher information matrix formalism, combining the results from these two experiments with those expected from the on-going CMB Planck experiment. The Planck CMB Fisher matrix contains seven parameters
 \begin{equation}
 \Omega_{\rm dm}h^2,\Omega_{\rm b} h^2,H_0, n_{\rm s},w, A_{\rm s},n_{\rm run}, 
 \end{equation}
 where $w$ is the dark energy equation of state, $A_{\rm s}$ is the amplitude of the primordial spectrum, $n_{\rm s}$ is the spectral index and $n_{\rm run}$ is the spectral index running. The galaxy survey Fisher matrix contains one additional parameter, the growth factor $\tilde{f}$, plus a galaxy bias which is redshift dependent. We have marginalized over the galaxy bias in each redshift bin.

BOSS-like survey parameters are the following: redshift coverage $0 < z < 0.7$,  $A_{\rm sky} = 10000$ deg$^2$ and a mean galaxy density of $2.66 \times 10^{-4}$.  For the EUCLID-like survey we assume a redshift coverage of $0.15 < z < 2$, a sky area of $f_{\rm sky} = 20000$ deg$^2$ and a mean galaxy density of $1.56 \times 10^{-3}$. We have verified that our errors are fully consistent with  those obtained by means of the Fisher routine provided by Ref.~\cite{White:2008jy}.

Firstly,  note from Fig.~\ref{fig:fs8} that, for the values of the
parameters chosen in this figure, the ${\tilde f} \sigma_8$ values for the
DMvel class I model of Eq.~(\ref{eq:maart}) are much lower than
current estimations. As carefully explored in
Ref.~\cite{CalderaCabral:2009ja}, negative values of $\Gamma$
(corresponding to positive $Q$) show a suppression of structure
growth due to a smaller amount of dark matter in the past. The
quantity ${\tilde f} \sigma_8$ will get depleted then, not only due to a lower
growth factor ${\tilde f}$, but also due to a lower $\sigma_8$. However,
current errors on $f \sigma_8$ measurements are large and do not rule
out these models at a high significance level, and other cosmological
parameter choices could differ in the ${\tilde f} \sigma_8$
predictions. Secondly, note that the predictions for the DMvel class
II model are very close to those of a non interacting model, as
expected from the results presented on dark matter velocities for this
model in  \S \ref{lineargrowthblah} and in \S \ref{sec:bulk}. Finally,
we can observe from Fig.~\ref{fig:fs8} that for the DEvel class II
model, huge ${\tilde f} \sigma_8$ values are predicted, especially at low
redshifts, when dark energy starts to be dominant and the effect of
the energy-momentum transfer starts to be important in the dark matter
differential velocity equation. Such an effect can not be obtained in the
two DMvel class I and class II models explored here because the Euler equation for the dark matter velocity is not modified in these interacting cosmologies. 
In the DEvel class II model, couplings $\xi<-0.4$ will be ruled out at more
than 3$\sigma$ with existing ${\tilde f} \sigma_8$ data.  

Accurate measurements of ${\tilde f}\sigma_8$ from future BAO surveys as BOSS or EUCLID may rule out significantly the form of couplings explored above, see the central and right panels of Fig.~\ref{fig:fs8}. 
For future surveys such as BOSS and EUCLID, the forecasted errors on the coupling $\xi$ for both the DMvel and DEvel class II models with $Q=\xi {\cal H} \rho_{\rm de}$ can be  as small as  $\Delta \xi =0.02$ and  $\Delta \xi <0.01$ respectively.
These forecasted errors should be considered as  optimistic, since in their derivation, we have assumed a perfect knowledge of the remaining cosmological parameters.

\subsection{Weak equivalence principle violation (WEPV) constraints} 
\label{sec:wepvnew}
Violation of the equivalence principle in the dark sector is a general feature of DEvel models. Regardless of the underlying physics responsible for this effect, it can modeled phenomenologically  by attributing to dark matter particles a ``fifth force'' specified by its potential.

\subsubsection{Coupled scalar field as an example of  DEvel $Q\propto \rho_{\rm dm}$ models}

Kesden and Kamionkowski (K$\&$K in the
following)~\cite{Kesden:2006zb,Kesden:2006vz}, analyzed the
consequences of WEP violation for  dark-matter on galactic scales,
focusing on dark-matter dominated satellite galaxies orbiting much
larger host galaxies. 
It was shown by running several N-body numerical simulations that in the
presence of a dark matter fifth force a pronounced asymmetry appears in the leading compared to the trailing tidal stream of the satellite galaxy.
In summary, they inferred  that a difference among dark matter and
baryonic accelerations larger than $10\%$ is severely disfavoured. 
They found that the leading-to-trailing ratio exceeds
$0.5$ for all simulations without a dark-matter force
and never exceeds $0.2$ for a dark-matter force with $4\%$ the
strength of gravity. From the number of simulations performed,
one can conclude that a difference between dark matter and baryon
accelerations larger than $4\%$ corresponds to at least a 2 $\sigma$
deviation in leading-to-trailing ratio. However a more quantitative statement
cannot be made as the way their simulations span the parameter space
for satellite interactions may not reproduce the same probability
distribution as real interactions.

To make this statement more
quantitative we review the case  DEvel class I models of the form of
Eq.~(\ref{kk}) considered in
e.g., \cite{Amendola2000a,Maccio:2003yk,Baldi,FriemanGradwohl,Wetterich95,Bean:2008ac,Koivisto:2005nr}. In these
models the effective gravitational potential between cold dark matter
particles is of the  Yukawa-type. In the limit of massless dark energy
interacting scalar field $\phi$, the strength of the gravitational interaction is
corrected by a $\beta^2$ contribution and reads:
\begin{equation}
G_{cc}=G_N(1+2 \beta^2)
\label{eq:gmod}
\end{equation}
where $G_N$ denotes the Newton's constant and $\beta$ parameterizes
 the dimensionless dark coupling of
Eq.~(\ref{kk}) as well as the strength of WEPV. 
It should be stressed  that in these coupled
quintessence models, the correction to the standard  gravitational
potential term  $k^2\Psi$ in the Euler equation for dark matter
velocity perturbations is due to a contribution $\propto 
-\beta k^2\delta\phi$ where $\delta
\phi$ is the scalar field perturbation. Indeed, it can be
shown~\cite{Amendola:2003wa}, that in these models $\beta\delta
 \phi\sim -\beta^2\Psi$, giving rise to a non negligible modification of gravity.
Using Eq.~(\ref{eq:gmod}), we see that the 10$\%$ limit  from  
K$\&$K  reduce to $|\beta|< 0.22$ in our notation.

The authors of Ref.\cite{Amendola:2003eq}  obtained, in the notation used here,  a 95\% bound of  $|\beta| \lesssim  0.11$ using CMB observations.
More recently, it has been argued~\cite{Peebles:2009th,Keselman:2009nx} 
that larger differences among dark matter and baryon accelerations, with an extra force as large as gravity, could be allowed if a screening length of $r_s =1$~Mpc/$h$ is assumed. This screening allow
the authors of Ref.~\cite{Peebles:2009th,Keselman:2009nx} to evade cosmological-scales constraints (CMB).  
The caveat to this argument is that CMB constraints apply to redshift 
$z\simeq 1100$ and
the parameter $\beta$ may vary with redshift. Ref. \cite{Sealfon} find
that large-scale structure observations  at $z < 0.2$ do not show any
indication for a typical scale where this Yukawa-type  modification
switches on, but  constrain the strength of the modification only
weakly;  \cite{dore} improve these bounds.

A possibly promising avenue to be explored from  forthcoming large,
SZ-selected  galaxy clusters surveys, is that of the baryon mass
fraction in galaxy clusters.  In fact, as shown in \cite{Baldi}, a
WEPV , with increasing coupling  reduces the clusters baryon fraction:
a $\beta=0.2$ reduces $f_{\rm b}$ in clusters  by 10\% with respect of
a $\Lambda$CDM model. This effect is mostly due to a mis-match of
velocities on large-scales not driven by
Eq.~(\ref{eq:gmod}). Tantalizingly,  this could alleviate the tension
between cosmological and clusters determinations of $f_{\rm b}$ from
current data, however to-date the error-bars are still large.
The results of \cite{Baldi} derive from N-Body
simulation for coupled models.  This type of analysis had already been
first performed in~\cite{Maccio:2003yk} and ~\cite{Baldi} claims to
obtain similar results on the enhancement of the bias in the nonlinear region
within and around massive halos. On a another hand,  let us mention that the
results of~\cite{Baldi} completely disagree with those
of~\cite{Maccio:2003yk} on the  increase or decrease of the inner halo
overdensity with respect to ΛCDM assuming the existence of a coupling
between the scalar field and dark matter. More
constraints have been derived on these type of coupled models using background
up to cosmological perturbations evolution (see
e.g.~\cite{Koivisto:2005nr,Bean:2008ac}), and can affect the late-time
Integrated Sachs-Wolfe (ISW) effect at large
scales~\cite{Xia:2009zzb}. See also~\cite{Baldi:2010vv} for
constraints on time
dependent coupling.   

Let us repeat  that on cosmological scales, a WEPV only in the dark
sector, would yield a mis-match between the baryon (and thus galaxy)
distribution and  the dark matter one \cite{Keselman:2009nx} which is
scale-dependent (see also \cite{Baldi}).   By comparing left (weak lensing)  and right (galaxy clustering) panel for fig. 2 of  \cite{dore} we conclude that there is no evidence for a large mis-match from present data, however the error-bars are still large. Forthcoming gravitational lensing surveys, combined with galaxy and Lyman alpha surveys should therefore constrain these models further. 

So far we have considered very specific models of the DEvel family for which a potential can be specified. The above consideration hold qualitatively also for the more phenomenologically-defined models as we show next. 

\subsubsection{DEvel  $Q\propto \rho_{\rm de}$ model }
As an example, we illustrate also the effect of from WEPV for the DEvel  class II toy-model considered along this paper and characterized by $Q_\nu= \xi  H \rho_{\rm de} u_{\nu}^{\rm de}$. In this DEvel class II model still the dark matter and baryon accelerations are clearly different, i.e. compare Eq.~(\ref{eq:thetames}) with $b=0$ to Eq.~(\ref{eq:thetab}). The results obtained in the following for this particular choice of coupling can be easily generalized to other choices of interaction.
At sub-horizon scales, in the Newtonian regime ($k\gg {\cal H}$), the baryon and dark matter accelerations are given by: 
\begin{eqnarray}
 \dot \theta_{\rm b}  & = & {\mathcal H} \dot \delta_{\rm b} + k^2 \Psi .\label{eq:dtb}\\
\dot \theta_{\rm dm}  & = & {\mathcal H}\left(1 +\xi \frac{\rho_{\rm de}}{\rho_{\rm dm}}\right)\left(\dot\delta_{\rm dm}+\xi{\cal H}
\delta_{\rm dm}\frac{\rho_{\rm de}}{\rho_{\rm dm}}\right)+ k^2 \Psi~,
\label{eq:dtdm}
\end{eqnarray}
where we have neglected $\delta_{\rm de}$ and  $\delta H$
contributions and $k^2 \Psi$ reads
\begin{equation}
k^2 \Psi=-\frac{3}{2} {\cal H}^2(\Omega_{\rm b} \delta_{\rm b} + \Omega_{\rm dm} \delta_{\rm dm})~.
  \label{eq:poisson}
\end{equation}
\begin{figure}[h!]
\begin{center}
\includegraphics[height=.4\textheight]{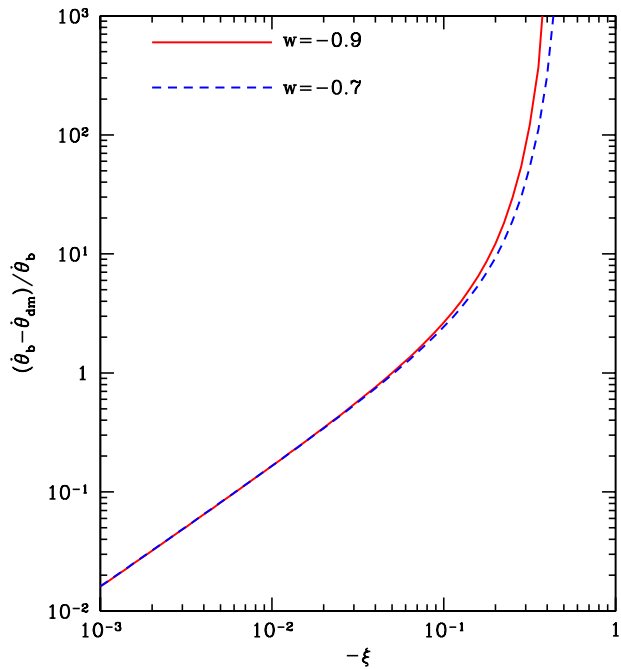}
  \caption{Relative dark matter-baryon acceleration $\frac{\dot \theta_{\rm b}- \dot
    \theta_{\rm dm}}{\dot \theta_{\rm b}}$ for $Q_\nu=\xi {\mathcal H} \rho_{\rm de} 
    u_{\nu}^{({\rm de})}/a$ using Eq.~(\ref{eq:dtb}) and~(\ref{eq:dtdm}). The cosmological parameters have been chosen to fit WMAP 5 year data, see text for more details. The red solid (blue short dashed) curve assumes an equation of state of the dark energy component $w=-0.9$ ($w=-0.7$).}
  \label{fig:WPr}
\end{center} 
\end{figure}
Figure~\ref{fig:WPr} shows the effect of the dark coupling on the
relative dark matter-baryon acceleration for the model illustrated here, 
for two different assumptions about the dark energy equation of state $w$. 
The cosmological parameters (as, for instance, $\Omega_{\rm dm} h^2$)
have been chosen accordingly to WMAP 5 year angular diameter distance
data~\cite{Komatsu:2008hk} as explained in section~\ref{lineargrowthblah}.

There is clearly a well defined relation between the strength of coupling and the mis-match between the dark matter and baryon acceleration that is robust to changes in $w$ and is well approximated by  a power law:

\begin{equation}
\ln\left(\frac{\dot{\theta}_{\rm b}-\dot{\theta}_{\rm dm}}{\dot{\theta}_{\rm b}}\right)=2.69+0.977\ln(-\xi)
\end{equation}
up to mis-matches of order of few $\times$ 100\%. This has been derived assuming linear theory and therefore quantitatively  strictly applies only to cosmological linear scales.

From the discussion above, our numerical results seems to indicate
that for DEvel class II models with $Q_\nu= \xi {\mathcal H} \rho_{\rm de}
u_{\nu}^{\rm de}/a$, couplings $\xi$ smaller than $\sim - 0.002$ would lead to a
$2\sigma$ sigma deviation in the leading-to-trailing ratio and couplings
smaller than  $\sim - 0.01$ are strongly disfavoured by K\& K numerical
analysis. Let us mention that those results are to be taken with care
given that no asymmetry in leading compare to the trailing of
e.g. Sagitarius dwarf galaxy is observed while K\& K numerical
best fit model tends to a leading-to-trailing ratio of 0.66 in the
absence of any dark matter fifth force

\subsection{Constraints from matter abundance in galaxy voids}
\label{sec:voids}
An uncontroversial (indisputable) feature of these models  arises from
the fact  dark energy is  smoothly, uniformly  distributed but dark
matter is not, and most of the volume of the Universe is occupied by
regions  under-dense of dark matter (voids). The coupling $Q \propto
\rho_{\rm de}$ characteristic of class II models requires that,
independently on the local dark matter density, a given amount of dark
matter per unit volume must turn into dark energy. Thus in the
underdense regions the coupling $Q \propto  \rho_{\rm de}$ must be
just an effective, coarse-grained  description: in fact, once locally
in an underdense region  all dark matter  has been transformed into
dark energy, the description adopted must  break down. In addition,
the uniform  depletion of dark matter   makes  the overdensity in
voids approach $\delta \sim -1$ very rapidly,  implying a prompt
breakdown  of linear perturbation theory.  While a detailed modeling
of this behavior may require numerical simulations, it is clear that
observations of  underdense regions  offer a promising avenue to
constrain such models. We consider now observations of voids properties to
constrain these models.

 Coupled dark energy-dark matter models lead to a dark matter depletion (or enhancement, depending on the sign and form of the interacting term). Most of this depletion/enhancement will take place from the background, which in turns translates in depleting/enhancing the voids in the large scale structure, which themselves occupy the most volume. Therefore, any measurement of the amount of matter in the voids with respect to the standard non-coupled model can help to constrain the value of the coupling. 

The matter content in voids has received recent attention (see e.g. Refs.~\cite{peebles,kyplin} and references therein) because it seems that the voids are more empty than expected from $\Lambda$CDM model predictions.  The work developed in Ref.~\cite{kyplin} summarizes very well our current knowledge on the occupancy of voids.    The local group, mostly our galaxy and Andromeda in mass content,  is about 20 Mpc away from the nearest cluster of galaxies, Virgo, and  sits in a relatively low-density region  e.g.,\cite{BinneyMerrifield}. Ref.~\cite{kyplin} argue that in the local volume, at a distance from us of about $10$~Mpc, there are too few small galaxies with circular velocities $V_c$ below $\sim 35$~km/s. They estimate a factor $10$ discrepancy with the $\Lambda$CDM model predictions. Other authors~\cite{peebles} have previously developed models to clear the voids of dark matter and thus quench the number of dark halos that could host these small galaxies. Of course, another explanation is that the $\Lambda$CDM model is correct and some galaxies are simply not visible \cite{jimdark,jimdark1}. Coupled models in which  a given amount of dark matter per unit volume  turns into dark energy, regardless of the local density  would  provide, therefore, a mechanism to clear dark matter (and galaxies) from the voids. On the other hand, the void phenomenon disfavors coupled models in which the dark matter  content per unit volume is enhanced. 
As an illustration, we use recent void results~\cite{kyplin} to derive
constraints on DMvel and DEvel class II models which will lead to a
depletion of the dark matter energy density of the universe, in
particular, the dark coupled model of Eqs.~(\ref{eq:ourm}) characterized by the
dimensionless coupling $\xi$. The latter cannot be too large and negative to empty the voids completely because, some galaxies with circular velocities $V_c$ below ~$35$~km/s are still observed in voids.

Before getting a precise estimate on the value of $\xi$, one can already impose a preliminary lower bound on the coupling parameter by requiring the universe as a whole not to be empty of matter in the voids today. Figure~\ref{fig:beth} shows the current cold dark matter energy density  in the universe as a function of the coupling for the dark sector interaction given by Eq.~(\ref{eq:ourm}) necessary to fit WMAP 5 year  data~\cite{Komatsu:2008hk}.   The  effect of the WMAP parameters uncertainties is illustrated by the two lines, chosen to vary along the parameter-degeneracy that most affect this figure.  This rather simple argument is indicating us that $\xi > -0.6$, see \S \ref{lineargrowthblah}. In what follows we will improve this lower limit on $\xi$. To proceed, we will assume that dark matter halos associated to galaxies with circular velocities  $\lesssim 35$~km/s are depleted by a factor of $10$ as argued in \cite{kyplin}. Note that at circular velocities of $\sim 50$~km/s there is no discrepancy with a $\Lambda$CDM model \cite{Trujillo}.  
We further argue that such a small galaxies and their associated halos  in low density regions form recently, when  dark energy  is important.  In fact   photo-ionization before recombination blows gas away from the halos of small mass. Only those small mass halos that collapse at $z < 1$ are able to retain their gas and thus form stars (e.g.  \cite{ferrara}  and references therein).
With this information we can compute the maximum  allowed  dark matter
depletion from the local void and translate it into a bound on
$\xi$. In fact  the velocity function  can be transformed into a mass
function using the spherical collapse (i.e., $35$~km/s corresponds to
$\sim 8\times 10^9M_{\odot}$). At these small masses the velocity
function and  mass functions (both baryonic Ref.~\cite{panter} and
dark matter) are well approximated by power laws. We thus modify the
$\Lambda$CDM-predicted power law for the local volume  mass function
to match the factor of 10 depletion below $35$~km/s, and still
matching the $\Lambda$CDM prediction at  $50$~km/s, concluding that
the total dark matter  mass in voids can at most be depleted by
20\%. This translates into a lower bound for the coupling of $\xi =
-0.2$, see Fig.~\ref{fig:beth}. This is a sharp lower limit, as it
relies on the results of Ref.~\cite{kyplin}, where error bars are way
much smaller than the size of the effect.  If we adopt a more
conservative point of view, arguing that  the matter is not actually
missing from the voids, but that star formation in small halos is
quenched, then the coupling $\xi$ will be limited to the range $-0.2 <
\xi < 0$. This constraint is more restrictive than those obtained in
Refs.~\cite{Gavela:2009cy,Honorez:2009xt}) from CMB, LSS and Supernova
Ia datasets. Therefore, future, more precise estimations of matter
abundance in voids could restrict the phenomenology of DEvel coupled
models in a stronger way than future CMB experiments as Planck or
EPIC, see Ref.~\cite{Martinelli:2010rt}. As our calculations suggests,
voids offer a promising avenue to the test dark coupling. To make our
argument more detailed and quantitative, theoretical modeling need to
be complemented by numerical simulations. 

\section{Conclusions}
\label{sec:concl}
In this paper we have shown that, although the number of interacting cosmologies that have been proposed in the literature is vast, it is always possible to classify all
existing   dark coupling models  in two broad families (DEvel and DMvel).  Within these families we examine two subclasses of models depending on whether the coupling scales like the density of dark matter ($\propto \rho_{\rm dm}$) or of dark energy ($\propto \rho_{\rm de}$).  At the background evolution level there are no differences between DEvel models and DMvel models:  the background evolution depends only on whether the coupling scales like the density of dark matter or of dark energy. However at the perturbation level the different classes show different phenomenologies which we have explored here.

\begin{figure}[h!]
\begin{center}
\hspace*{-1.cm}
\includegraphics[height=.5\textheight]{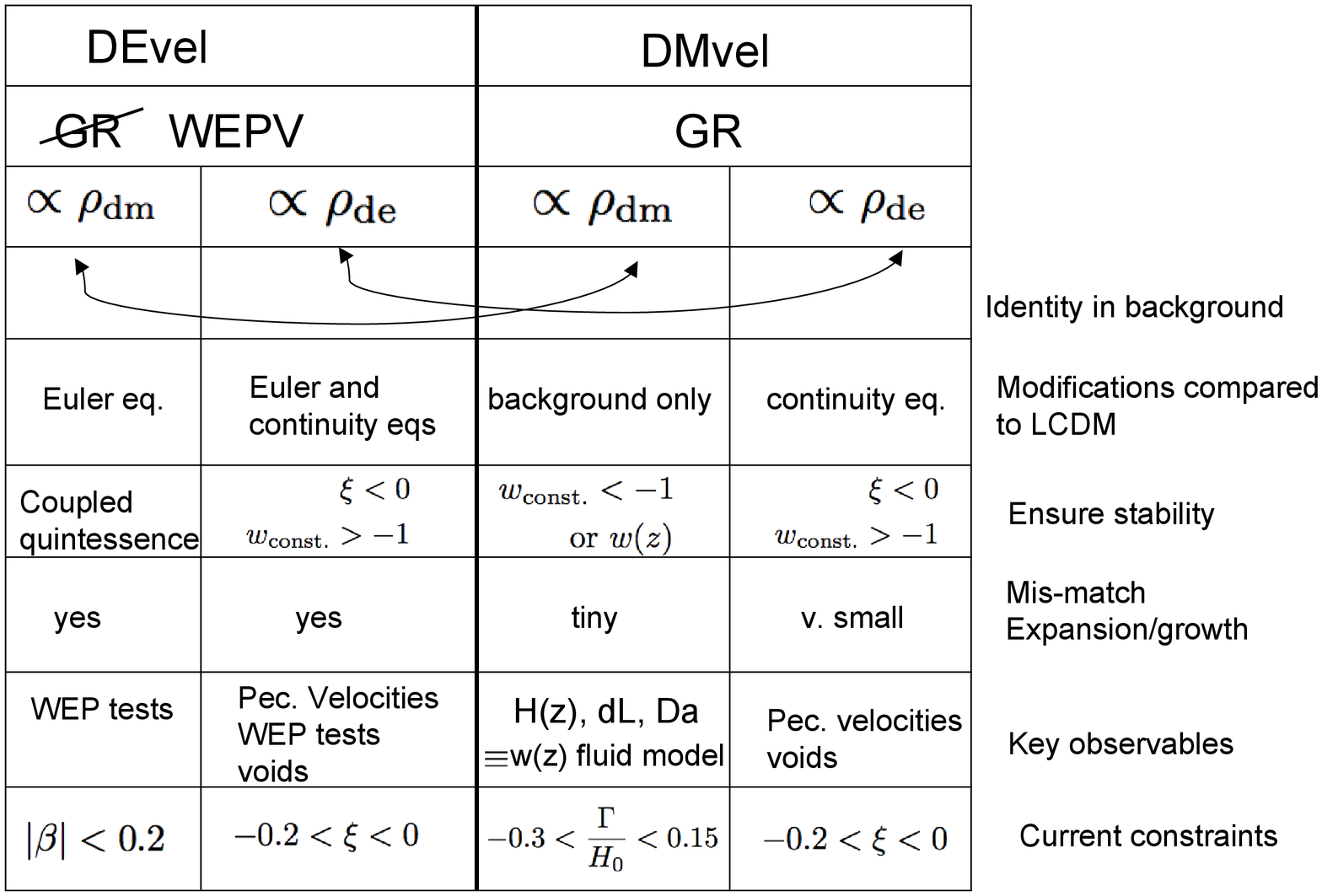}
 \caption{Summary of classification, phenomenology and current constraints on dark coupling models. $d_L$ denotes the luminosity distance  as obtained from e.g. Supernovae Ia observations and $D_a$ denotes the angular diamater distance as measured e.g., by angular BAO.}
 \label{table}
\end{center}
\end{figure}

GR predicts an exact relation between $H(z)$ and the growth of structure for non-interacting dark fluids.  Interactions modify this relation, and so can appear like a modification of gravity when the  growth of structure and expansion history are compared precisely.  We have shown that DEvel models, where there is no momentum transfer to the dark energy rest frame, induce a ``fifth force'' on the dark matter proportional to its peculiar velocity, violating the weak equivalence principle and deviating substantially from the uncoupled case in the growth of structure.  However, for DMvel models, deviations from uncoupled growth are very small.  Finally, DMvel models $\propto \rho_{\rm dm}$ are effectively indistinguishable from minimally coupled dark energy models   with time evolving equation of state parameter.  
 Therefore, if  as a result of a forthcoming experiment, the  measured expansion history and growth  of structure are in agreement, this observation can be used to constrain deviations from GR, as commonly accepted in the literature. The conversely, however,  is not necessarily true: a mis-match between expansion and growth could,   indicate deviations from GR but could also,  in principle indicate a dark coupling with GR unchanged. For the DMvel coupled models we considered, however, this effect is small.   This is summarized in Fig. \ref{table}.

We have analyzed how low redshift and near universe probes  could be used to constrain  coupled dark matter-dark energy scenarios. We have considered mis-match between high-redshift (Cosmic microwave background--CMB) constraints and low redshift measurements of the background quantities such as the Hubble parameter.  We have also considered tests such as the skewness, which may be promising only for DMvel $\&$ DEvel $\propto \rho_{\rm de}$ models; in these models deviations in the skewness for the dark matter distribution could reach the $10\%$ level. It is however not clear whether this effect could be measured or it would be degenerate with the dark matter-galaxy bias. For the baryon distribution,  (and for DMvel and DEvel $\propto \rho_{\rm dm}$ models matching CMB constraints) the deviation barely reaches $ 1\%$.
We have also revisited velocity-related probes, such as  local bulk flows and redshift space distortions. Present data  are not very constraining for DMvel models, but future redshift-space distortions measurements from  on-going and proposed galaxy surveys  have the potential to tighten significantly the allowed coupling window for both DMvel and DEvel models.
Finally, we have shown that in the context of DMvel $\&$ DEvel $\propto \rho_{\rm de}$
models with negative coupling, the matter abundance in voids can provide
a method to constrain interacting models efficiently.
This test only depends on the background evolution and can be applied to any
interacting model depleting or enhancing the dark matter abundance
homogeneously in time.

For each of the dark coupling  classes we have considered viable and stable models that fit CMB observations, and confronted them with present  low redshift probes. We have  also quantified the potential of future data.
The current status on the constraints on  the viable dark coupling models is summarize in Fig. \ref{table}. The   reported constraints on DEvel $\propto \rho_{\rm dm}$, DEvel $\propto \rho_{\rm de}$, DMvel $\propto \rho_{\rm dm}$ and DMvel $\propto \rho_{\rm de}$ come respectively from weak equivalent principle violations of galactic scales, voids, redshift-space distortions and voids.

The combination of  precision tests of expansion history  (BAO,
$H(z)$, supernovae, matter content in voids) and growth of structure
tests (peculiar velocities and weak lensing) together with weak
equivalence principle  tests on astronomical scales (galaxies and
satellites dynamics)  is the key to explore or constrain a possible
coupling in the dark sector.


\section*{Acknowledgments}

 L.~L.~H was partially supported by CICYT through
the project FPA2006-05423, by CAM through the project HEPHACOS, P-ESP-00346,  by the PAU (Physics
of the accelerating universe) Consolider Ingenio 2010, by the
F.N.R.S. and  the I.I.S.N..  BAR was supported by OISE/0530095. O.~M. work is supported by the MICINN Ram\'on y Cajal contract, AYA2008-03531 and CSD2007-00060. LV is supported by FP7-PEOPLE-2007-4-3  IRG n 202182, FP7 IDEAS-Phys.LSS.240117. LV and RJ are supported by  MICINN grant AYA2008-03531. 
\appendix
\section{Background evolution in models with $Q=\xi \cal{H} \rho_{\rm de, dm}$}
For DEvel or DMvel models with $Q=\xi \cal{H} \rho_{\rm de}$, the Hubble expansion rate function as a function of the redshift reads~\cite{Gavela:2009cy} 
\begin{equation}
H(z)=H_0\sqrt{ \Omega_{\rm dm}^{(0)} (1+z)^{3} +
      \Omega_{\rm de}^{(0)}\frac{\xi}{3w^{\rm eff}_{\rm de}}(1- (1+z)^{3w^{\rm eff}_{\rm de}})  (1+z)^{3}
  + \Omega_{\rm de}^{(0)}(1+z)^{3(1+w^{\rm eff}_{\rm de})}}~,
\label{eq:hz}
\end{equation}
and the redshift dependent equation of state $\tilde w(z)$ that one
would reconstruct from Eq.~(\ref{eq:wH}) reads~\cite{Gavela:2009cy}
 \begin{equation}
   \label{eq:wzdelta}
\tilde w(z)=\frac{w}{1-\frac{\xi}{3w^{eff}_{de}}(1- (1+z)^{-3w^{eff}_{de}})}\,,
 \end{equation}
where $w_{\rm de}^{\rm eff}= w+\frac {\xi}{3}$.
If $Q=\xi {\cal H}\rho_{\rm dm}$ then:

\begin{equation}
H(z)=H_0\sqrt{ \Omega_{\rm dm}^{(0)} (1+z)^{3(1+w_{\rm dm}^{\rm eff})} +
     \Omega_{\rm dm}^{(0)}\frac{(1+z)^3}{1-\frac{w}{w^{\rm eff}_{\rm dm}}}((1+z)^{3w}-
(1+z)^{3w^{\rm eff}_{\rm dm}}) + \Omega_{\rm de}^{(0)}(1+z)^{3(1+w)}}~,
\label{eq:hz2}
\end{equation}
where $w_{\rm dm}^{\rm eff}=-\frac {\xi}{3}$.The redshift dependemt equation of state $\tilde w(z)$ that one
would reconstruct from Eq.~(\ref{eq:wH}), reads, at small redshifts~\cite{Gavela:2009cy}  
\begin{equation}
\tilde w(z)=w \left(1+\xi \frac{ \Omega_{\rm dm}^{(0)}}{ \Omega_{\rm de}^{(0)}} z \right)~.
\end{equation}
\section{Cosmological parameters according to WMAP 5 constraint}
\label{sec:cosm-param-accord}
Assuming a flat universe and perfect measurements of $\Omega^{(0)}_{\rm dm} h^2$,
$\Omega^{(0)}_{\rm b} h^2$, and the angular diameter distance to the last
scattering surface from CMB observations, the amplitude of $\xi$ is
degenerate with the physical energy density in dark matter
today $\Omega^{(0)}_{\rm dm} h^2$. Table \ref{tab:tab1} illustrates for a selected range of couplings this effect on some of the cosmological parameters that have been used in the numerical calculations presented here.
\begin{table}
\centering
\begin{tabular} {c c c}
\hline \hline
$\xi$ &$\Omega^{0}_{\rm dm}h^2$&$H_0$(km/s/Mpc)\\
\hline \hline
 0&0.1099&68.29\\
-0.1&0.097&69.48\\
-0.2&0.083&70.68\\
-0.3&0.067&71.90\\
-0.4&0.050&73.13\\
-0.5&0.032&74.37\\
-0.6&0.012&75.63\\
\hline \hline
\end{tabular}
\label{tab:tab1}
\caption{Current values of the dark matter energy density and the Hubble parameter necessaries to fit WMAP 5 year angular diameter distance data as a function of the coupling $\xi$. For numerical purposes, each time that the coupling is varied, the present day values of the cosmological parameters are choosing accordingly to the results of the fit illustrated here.}
\end{table}

\bibliographystyle{hunsrt} 

\providecommand{\href}[2]{#2}\begingroup\raggedright\endgroup

%
%
\end{document}